\documentclass{article}

\usepackage{multirow}
\usepackage{arxiv}

\usepackage[utf8]{inputenc} % allow utf-8 input
\usepackage[T1]{fontenc}    % use 8-bit T1 fonts
\usepackage{hyperref}       % hyperlinks
\usepackage{url}            % simple URL typesetting
\usepackage{booktabs}       % professional-quality tables
\usepackage{amsfonts}       % blackboard math symbols
\usepackage{nicefrac}       % compact symbols for 1/2, etc.
\usepackage{microtype}      % microtypography
\usepackage{lipsum}
\usepackage{graphicx}
\graphicspath{ {./images/} }
\usepackage{xcolor}
\usepackage[most]{tcolorbox}

\title{Synergy between Machine/Deep Learning and \\ Software Engineering: How Far Are We?}

\author{
 Simin Wang \\
  Department of Computer Science\\
  Southern Methodist University\\
  Dallas, TX 75275 \\
  \texttt{siminw@smu.edu} \\
   \And
 Liguo Huang \\
  Department of Computer Science\\
  Southern Methodist University\\
  Dallas, TX 75275 \\
  \texttt{lghuang@lyle.smu.edu} \\
  \And
 Jidong Ge \\
  State Key Laboratory of Novel Software Technology\\
  Nanjing University\\
  Nanjing, China \\
  \texttt{gjd@nju.edu.cn} \\
  \And
 Tengfei Zhang \\
  Software Institute\\
  Nanjing University\\
  Nanjing, China \\
  \texttt{terryzhang1009@foxmail.com} \\
  \And
 Haitao Feng \\
  Software Institute\\
  Nanjing University\\
  Nanjing, China \\
  \texttt{fenghaitaofht@gmail.com} \\
  \And
 Ming Li \\
  School of Artificial Intelligence\\
  Nanjing University\\
  Nanjing, China \\
  \texttt{lim@lamda.nju.edu.cn} \\
  \And
 He Zhang \\
  State Key Laboratory of Novel Software Technology\\
  Nanjing University\\
  Nanjing, China \\
  \texttt{hezhang@nju.edu.cn} \\
  \And
 Vincent Ng \\
  Human Language Technology Research Institute\\
  University of Texas at Dallas\\
  Richardson, TX 75083 \\
  \texttt{vince@hlt.utdallas.edu} \\
}

\begin{document}
\maketitle

\begin{abstract}
Since 2009, the deep learning revolution, which was triggered by the introduction of ImageNet, has stimulated the synergy between Machine Learning (ML)/Deep Learning (DL) and Software Engineering (SE). Meanwhile, critical reviews have emerged that suggest that ML/DL should be used cautiously. To improve the quality (especially the applicability and generalizability) of ML/DL-related SE studies, and to stimulate and enhance future collaborations between SE/AI researchers and industry practitioners, we conducted a 10-year Systematic Literature Review (SLR) on 906 ML/DL-related SE papers published between 2009 and 2018. Our trend analysis demonstrated the mutual impacts that ML/DL and SE have had on each other. At the same time, however, we also observed a paucity of replicable and reproducible ML/DL-related SE studies and identified five factors that influence their replicability and reproducibility. To improve the applicability and generalizability of research results, we analyzed what ingredients in a study would facilitate an understanding of why a ML/DL technique was selected for a specific SE problem. In addition, we identified the unique trends of impacts of DL models on SE tasks, as well as five unique challenges that needed to be met in order to better leverage DL to improve the productivity of SE tasks. Finally, we outlined a road-map that we believe can facilitate the transfer of ML/DL-based SE research results into real-world industry practices.
\end{abstract}

% keywords can be removed

\section{Introduction}
\begin{quote}
\textit{We as an industry have not yet built enough of these AI systems to fully understand how they might impact the software engineering process, as they most certainly will}. --- Grady Booch \cite{Booch18} 
\end{quote}

Software Engineering (SE), as noted by UML's co-orginator Grady Booch, 
%one of the UML's originators, offered his perspective that Software Engineering (SE) 
is undergoing a dramatic change fused by big data and recent advances in Artificial Intelligence (AI). Despite the two AI winters in the last century \cite{wiki:xxx}, tremendous achievements in Machine Learning (ML), especially Deep Learning (DL), in the 2010s blew away the haze and started the season. Among these, ImageNet, a large-scale ontology of images, was created by Fei-Fei Li's group in 2009, providing training and benchmark data for image classification algorithms, such as Convolutional Neural Networks (CNNs) \cite{Deng09}. At almost the same time, a leap in Nvidia's graphics processing units (GPUs) significantly reduced the computation time of DL algorithms. Both milestone accomplishments became the catalyst for the ML/DL boom in all applications, and SE has certainly become one of the beneficiaries. Xie et al. \cite{Xie2009} conducted an empirical study on data mining involving ML methods for SE in 2009, which described several challenges of mining SE data and proposed various algorithms to effectively mine sequences, graphs and text from such data. The study projected to see the expansion of the scope of SE tasks that could benefit from data mining and ML methods as well as the range of SE data that can be mined and learned from.

Over the past decade, SE researchers have been showing a great enthusiasm for exploring and applying ML/DL techniques to diverse tasks for various SE activities\footnote{SE activities refer to the fundamental activities in all software development processes, including requirements engineering, design and modeling, implementation, testing, defect analysis, maintenance and evolution, and project management.}. ML and DL accelerate the contemporary software development lifecycle by analyzing huge amounts of data produced and consumed throughout the process. The results of these insights are driving changes in software development and evolution. For instance, in requirements engineering, ML is employed to identify relevant requirements in software specifications \cite{Knauss12} and generate requirement traceability links between textual requirements artifacts \cite{Sultanov13}. In software design and modeling, various supervised classifiers are used to detect the presence of architectural tactics in code, tracing them to requirements, and visualizing them in a way that helps a developer to understand the underlying design decisions \cite{7270338}. During implementation and testing, ML/DL aids developers with a variety of SE tasks, such as code generation and completion, code optimization, API recommendation, and test case generation. For defect analysis, almost all types of ML/DL models are leveraged to classify defects \cite{Huang2015}, predict defective software artifacts \cite{Yan17}, locate the faulty program elements that are responsible for the failure \cite{Le13} and generate a prioritized list of relevant bug-fix suggestions \cite{Jeffrey09}. To facilitate maintenance and evolution tasks, researchers have mined various repositories using machine learning algorithms such as Support Vector Machine (SVM), Naive Bayes and Logistic Regression to extract high-quality aligned data \cite{Yin18}, identify API issue-related sentences \cite{Ahasanuzzaman18}, and classify user reviews from app stores \cite{GENCNAYEBI2017207}.

With the rapid growth of ML/DL techniques, it appears that ``Spring is coming'' for the SE community. Recently, controversial studies have emerged that warn us to use ML/DL cautiously and not overlook the exploration of simple methods. Liu et al. \cite{Liu:2018:NCM:3238147.3238190} proposed a nearest neighbor algorithm that does not require any training to generate short commit messages, which is not only much faster and simpler but also performs better than the Neural Machine Translation approach by Jiang et al. \cite{Jiang:2017:AGC:3155562.3155583}. Xu et al. \cite{Xu:2016:PSL:2970276.2970357,Xu:2018:PRS:3239235.3240503} and Fu and Menzies \cite{Fu:2017:EOH:3106237.3106256,Menzies18} debated the effectiveness of DL vs.\ SVM in predicting the relatedness between Stack Overflow knowledge units. Eventually, they agreed in part that while SVM based approaches offer slightly better performance, they are slower than DL-based approaches on larger datasets.

To stimulate and enhance the future collaborations between the SE and AI communities, 
%for the next decade, 
we conducted a Systematic Literature Review (SLR) to comprehensively evaluate the impacts that ML/DL and SE have on each other. Our goal is to raise sufficient awareness about the challenges and problematic situations in ML/DL-related SE studies. Specifically, this study makes the following contributions:

\begin{itemize}
  \item To the best of our knowledge, we are the first to carry out a comprehensive SLR on 906 papers published in the past decade, which either employ ML/DL techniques on SE tasks or tailor SE methods/practices to the development of ML/DL systems\footnote{In this paper, ML/DL systems refer to software frameworks, tools, or libraries that provide ML/DL functionalities, or software systems that have ML models or DNNs as their core with extra software encapsulation.}.
  \item We demonstrated the impacts that emerging ML/DL techniques have on SE tasks by investigating the breadth and depth of the changes that ML/DL techniques have brought to SE tasks. In addition, we discussed the challenges of applying SE practices to the development of ML/DL systems, typically for testing and debugging tasks. 
  \item We identified five factors that influence the replicability and reproducibility of ML/DL-based studies in SE. These factors can be assessed to improve the quality of future research.
  \item By categorizing the rationales behind the selection of ML/DL techniques into three types, we recognized that the rationales provided in our identified studies are often insufficient. Consequently, we identified three indispensable ingredients in the studies that would facilitate an understanding of why a ML/DL technique was selected to tackle a specific SE problem.
  \item We identified the unique trends of impacts of DL models on SE tasks, as well as five unique challenges that need to be met to better leverage DL to improve the productivity of SE tasks.
  \item Based on these synthesized findings, we created a road-map that could facilitate the transfer of ML/DL-related SE research results into real-world industry practices.
\end{itemize}

\section{Machine/Deep Learning}
The origin of ML can be traced back to 1959 by Arthur Samuel \cite{Samuel1988}. A widely quoted, more formal definition of ML was proposed by Tom M.\ Mitchell: % defined ML more formally as 
``a computer program is said to learn from experience E with respect to some class of tasks T and performance measure P if its performance at tasks in T, as measured by P, improves with experience E'' \cite{mitchell1997machine}. ML comprises four major learning paradigms, namely \textit{supervised learning (SL)}, \textit{unsupervised learning (UL)}, \textit{semi-supervised learning (SSL)} and \textit{reinforcement learning (RL)}. In \textit{SL}, the training data 
%of applications 
comprises examples (represented in the form of vectors) %of the input vectors 
along with their %corresponding 
target values
%vectors 
and its aim is to assign each input vector one of a finite number of discrete categories (classification) or a real value (regression). \textit{UL} is often used to discover groups of similar examples within the data (clustering), where none of its training examples is labeled with %while its training data consists of a set of input vectors x without 
the corresponding target value \cite{bishop2006pattern}. \textit{SSL} presents a challenging learning setting where models are constructed from %incomplete 
training data that is typically composed of a small number of labeled instances and a large number of unlabeled instances.
%where a portion of the sample input does not have target values. 
Finally, \textit{RL} is concerned with the problem of finding suitable actions to take in a given situation in order to maximize a reward by interacting with the surrounding environment \cite{bishop2006pattern}.

Deep learning is a branch of ML built on multiple levels of representation obtained by composing simple but non-linear modules. Starting with the raw input, each module transforms the representation at one level into a representation at a higher, more abstract level \cite{lecun2015deep}. Compared to conventional ML algorithms, DL methods are very effective at exploring high-dimensional data \cite{Fu:2017:EOH:3106237.3106256}.

\section{Research Method}
This study was initiated in the middle of 2018. Following the approach proposed by Kitchenham and Charters \cite{kitchenham2007guidelines} and Zhang et al.~\cite{ZHANG2011625}, a SLR was conducted based on a rigorous research strategy. The leading author of this study is a PhD candidate whose research interest lies in employing ML/DL techniques to facilitate the exploration of challenging SE tasks. The remaining co-authors, of which two are our supervisors, have long-term experiences with either SE or ML/DL. This section describes the research methodology used for conducting this study.

\subsection{Research Questions and Motivations} 
This empirical study aims to answer the following research questions:
\begin{itemize}
\setlength{\itemindent}{-1em}
\item[] \textbf{RQ1.} \textit{What are the trends of impacts of emerging ML/DL techniques on SE tasks, as well as those of SE methods on ML/DL system development from 2009 to 2018?}
\item[] \textbf{RQ2.} \textit{ What are the factors that influence the replicability and reproducibility of ML/DL-based studies in SE?}
\item[] \textbf{RQ3.} \textit{ What ingredients in a study would facilitate the understanding of why a ML/DL technique is selected for a specific SE task?}
\item[] \textbf{RQ4.} \textit{ What are the unique impacts and challenges for research in leveraging DL to solve SE problems?}
\end{itemize}

\textit{RQ1} attempts to summarize the emerging changes (accomplishments and deficiencies) we discovered as part of %the synergy between ML/DL and SE by analyzing 
our analysis of the trend of ML/DL-related SE studies. 
%To enhance the synergy between ML/DL and SE, the findings of 
Through \textit{RQ2}, \textit{RQ3} and \textit{RQ4}, we hope to shed light on issues concerning how to improve the applicability and generalizability of ML/DL-related SE studies. 
Specifically, \textit{RQ2} attempts to understand how to make ML/DL applications to SE more replicable and reproducible. 
%when applying and generalizing to other SE tasks. 
\textit{RQ3} investigates how detailed the rationales provided by SE researchers are with respect to the choice of a ML/DL technique used for a particular SE task and how to improve the description of these rationales. %Due to the unique contributions in learning and the challenges to be employed for SE tasks, 
Given the recent surge of interest in applying DL to SE tasks, \textit{RQ4} examines the unique challenges of employing DL in SE studies and the unique impacts of DL applications on SE, as well as promising future directions for DL research in SE.

\begin{table}[t]
\footnotesize
  \caption{Publication venues for manual search.}
  \renewcommand{\arraystretch}{1.3}
  \label{tab:Venue}
  \begin{center}
  \begin{tabular}{l p{10cm}}
    \hline 
    \textbf{Acronym} & \textbf{Venues}\\
    \hline
        ICSE & International Conference on Software Engineering\\
        ASE & International Conference on Automated Software Engineering\\
        ESEC/FSE & European Software Engineering Conference and International Symposium on Foundations of Software Engineering\\
        ICSME & International Conference on Software Maintenance and Evolution\\
        ICPC & International Conference on Program Comprehension\\
        ESEM & Symposium on Empirical Software Engineering and Measurement\\
        RE & Requirements Engineering Conference\\
        ISSTA & International Symposium on Testing and Analysis\\
        MSR & Working Conference on Mining Software Repositories;\\
        SANER & International Conference on Software Analysis, Evolution and Reengineering\\
        ESE & Empirical Software Engineering\\
        TSE & IEEE Transactions on Software Engineering\\
        TOSEM & ACM Transactions on Software Engineering and Methodology\\
        JSEP & Journal of Software: Evolution and Process\\
        JSS & Journal of Systems and Software\\
        IST & Information \& Software Technology\\
        AAAI & AAAI Conference on Artificial Intelligence\\
        IJCAI & International Joint Conference on Artificial Intelligence\\
        ACL & Meeting of the Association for Computational Linguistics\\
        ICML & International Conference on Machine Learning\\
        AI & Artificial Intelligence (Journal)\\
        JMLR & Journal of Machine Learning Research\\
        EMNLP & Empirical Methods in Natural Language Processing\\
        CoNLL & Computational Natural Language Learning\\
  \hline
\end{tabular}
\end{center}
\end{table}

\begin{table}[t]
\footnotesize
  \renewcommand{\arraystretch}{1.3}
  \caption{Study inclusion and exclusion criteria.}
  \label{tab:InEx}
  \begin{center}
  \begin{tabular}{p{12cm}}
    \hline
    \textbf{Inclusion criteria}\\
    \hline
    \vspace{-3mm}
    \begin{enumerate}
        \item The paper claims that a ML/DL technique is used
        \item The paper claims that the study involves an SE task or or one or more topics covered in the field of SE \cite{SEGlossary90}
        \item The full text of the paper is accessible
    \end{enumerate}
    \vspace{-3mm}\\
    \hline
    \textbf{Exclusion criteria}\\
    \hline
    \vspace{-3mm}
    \begin{enumerate}
        \item The number of pages is less than 8
        \item The authors claim that their study is a SLR, review or survey
        \item Short papers, tool demos and editorials
        \item The paper was published in a workshop or a doctoral symposium
        \item The paper is a grey publication, e.g., a technical report, a thesis
    \end{enumerate} 
    \vspace{-3mm}\\
  \hline
\end{tabular}
\end{center}
\end{table}

\subsection{Search Strategy}
In 2009, the deep learning revolution, which was triggered by the introduction of ImageNet, has transformed AI research in both academia and industry \cite{Fortune16}, and an early adoption of data mining and ML techniques in SE research was pioneered by Xie et al. \cite{Xie2009}. Therefore, we set 2009 as the starting year for our search of publications when preparing a 10-year review that spans the period from 2009 to 2018. Using the `Quasi-Gold Standard' method \cite{ZHANG2011625}, we integrated manual and automated search strategies in a relatively rigorous process. To begin with, we chose 16 top SE (ICSE, ASE, ESEC/FSE, ICSME, ICPC, RE, ESEM, ISSTA, MSR, SANER, TSE, TOSEM, ESE, IST, JSS, JSEP) and 8 AI (AAAI, IJCAI, ACL, ICML, AI, JMLR, EMNLP, CoNLL) conferences and journals that have published papers addressing the synergy between SE and ML/DL as the manual search venues, as shown in Table \ref{tab:Venue}. Correspondingly the follow-up automated search venues are IEEE Xplore, ACM Digital Library and SCOPUS. Then, through a careful evaluation \cite{ZHANG2011625}, we refined and finalized the search string for automated search as follows:

``\textit{(`machine learn*' OR `deep learning' OR `neural network?' OR `reinforcement learning' OR `unsupervised learn*' OR `supervised learn*') AND (`software engineering' OR (software AND defect) OR `software requirement?' OR `software design' OR `software test*' OR `software maintenance' OR `source code' OR `project management' OR `software develop*')}''\footnote{An asterisk (*) in a search term is used to match zero or more characters, and a question mark (?) is used to match a single character.}

\begin{figure}[t]
  \centering
  \includegraphics[scale = 0.9]{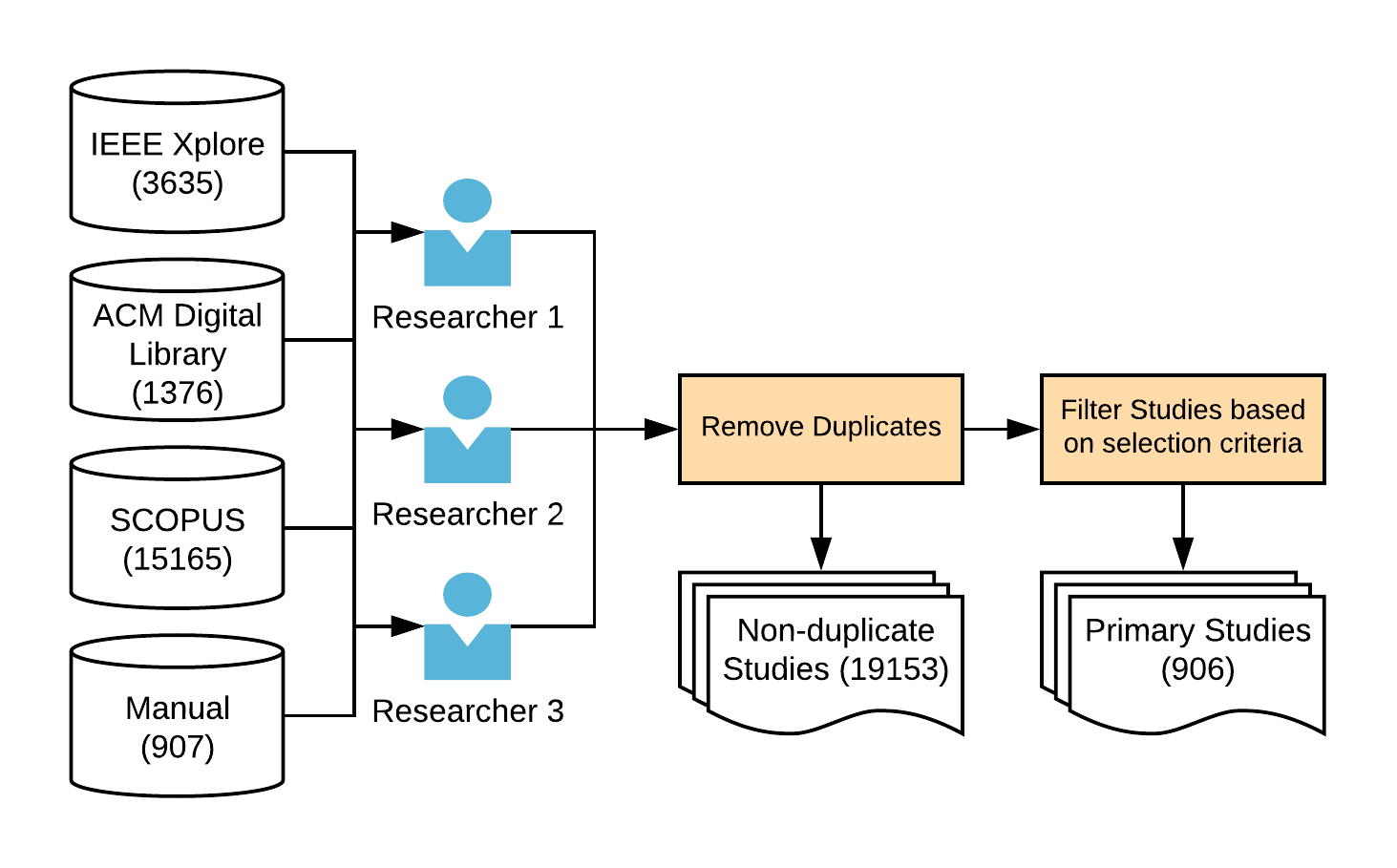}
  \caption{Study Selection Process.}
  \label{Study Selection}
\end{figure}

\subsection{Study Selection}
Once we retrieved the studies that was deemed potentially relevant based on our search strategy, an assessment of their actual relevance according to the inclusion and exclusion criteria in Table \ref{tab:InEx} was executed in order to select the primary studies that provide direct evidence about the research questions. The search and selection process took nearly seven months.

The selection procedure is divided into three phases (Figure \ref{Study Selection}): identification, deduplication and filtering. Initially, we retrieved 21,083 papers from three digital libraries and manual search, and discarded duplicates by inspecting titles and authors. This reduced the total number of papers to 19,153. Before moving to the final filtering, a pilot was conducted in order to establish a homogeneous interpretation of the selection criteria among three researchers \cite{Unterkalmsteiner12}. We randomly selected 30 publications from our collection and assessed them individually by full-text reading. If the Fleiss' Kappa value did not reach \textit{almost perfect agreement} according to Landis and Koch \cite{landis1977measurement}, which was our acceptable level for the final selection, the research team would hold a discussion, strive to reach a consensus and repeat the process. After applying the inclusion/exclusion criteria, 906 papers remained as primary studies and were used for data extraction. In some cases, our supervisors, the two domain experts in SE and ML/DL, gave their advice on ``Hard to Determine'' publications.

\begin{table}[t]
\footnotesize
  \caption{Data extraction.}
  \renewcommand{\arraystretch}{1.3}
  \label{tab:DE}
  \begin{center}
  \begin{tabular}{l p{10cm}}
    \hline 
    \textbf{RQ} & \textbf{Data item}\\
    \hline
        1 & Published year\\
        1,2,3,4 & The category of SE task \\
        1 & The SE activity to which the SE task belongs \\
        1,2,3,4 & The adopted ML/DL techniques\\
        2,3 & Whether and what technique comparisons are used\\
        2 & Whether the training and test data is published\\
        2 & Whether the source code is published\\
        2 & Whether the dataset is from industry, an open source or a student collection\\
        2 & Whether and how data preprocessing techniques are used\\
        2 & Whether and how hyper-parameter optimization techniques are used\\
        4 & The size of the data before and after preprocessing\\
        4 & Whether the study reports the runtime of the ML/DL model \\
        2,3 & Whether and how the authors describe the rationale behind techniques selection\\
        2,3 & Whether the study has an error analysis\\
        2 & The number of times a study has been successfully replicated or reproduced\\
  \hline
\end{tabular}
\end{center}
\end{table}

\subsection{Data Extraction}
Table \ref{tab:DE} presents the data items extracted from the primary studies, where the column `RQ' shows the related research questions to be answered by the extracted data items on the right.

Due to the large number of studies and time constraints, we distributed the workload to each researcher as follows. Each researcher was given 2/3 of all the studies in order to guarantee that all primary papers were assessed by at least two researchers \cite{kitchenham2007guidelines}. All the information was recorded in spreadsheets, the data extraction form. 
%Similar to the primary study selection, 
Another pilot of 20 randomly selected publications was performed by all three researchers to check data extraction consistency, during which the extracted data was cross-checked and any disagreement was resolved by discussion or expert advice. It is sometimes difficult to extract the rationales behind the selection of ML/DL techniques (i.e., identifying the discussion of the suitability and advantages of the selected methods and why the selected model works) as they are often scattered in the paper. In these cases, we examined three places where the rationales are most likely explained, including motivation examples, model design, and background. Then, we carefully analyzed and elicited the potential rationales. Some rework was needed to consolidate the extracted data. The entire data extraction phase took around six months.

\subsection{Study Quality Assessment}
To assess the quality of the primary studies, a quality checklist was designed (see Table \ref{tab:QA}) with three different choices: \textit{Yes}, \textit{Partially}, and \textit{No}. We answered the questions in the checklist and presented the distribution of papers over three choices.

\begin{table}[t]
\footnotesize
  \renewcommand{\arraystretch}{1.3}
  \caption{Study quality assessment.}
  \label{tab:QA}
  \begin{center}
  \begin{tabular}{p{1cm} p{9cm} p{1cm} p{1.5cm} p{1cm}}
    \hline
    \textbf{ID} & Quality assessment question & Yes & Partially & No\\
    \hline
        QA1 & Are the aim and motivation of the research described clearly? & 889 & 17 & 0\\
        QA2 & Is the presented methodology clearly explained? & 843 & 59 & 4\\
        QA3 & Are the research questions clearly defined and sufficiently explained? & 532 & 40 & 334\\
        QA4 & Is the contribution of the research clearly stated? & 605 & 291 & 10\\
        QA5 & Is the presented methodology evaluated by enough data? & 724 & 160 & 22\\
  \hline
\end{tabular}
\end{center}
\end{table}

With QA1 and QA2 we assessed if the authors of a study clearly stated the aims, the motivations and the methodologies of their research. For almost all publications these questions could be answered at least \textit{Partially}. For QA3, the answer is \textit{Yes} if the research questions were explicitly defined and addressed with details in the result analysis, and \textit{No} if research questions were not defined. According to the statistics in Table \ref{tab:QA}, this question had a polarization of answers. For QA4, the answer is \textit{Yes} if the research contributions were clearly presented. This question received positive answers for all but 10 studies. QA5 was evaluated based on the size of the training set used in ML/DL applications: \textit{Yes} if the data size $\geq 1,000 $, \textit{Partially} if the data size $< 1,000$, and \textit{No} if the data size is not provided. The result of QA5 shows that 724 (80\%) of 906 studies were evaluated on datasets with size greater than 1,000.

\subsection{Data Synthesis}
We applied both quantitative and qualitative methods to collate and summarize the results of the included primary studies. For \textit{RQ1}, the trend and distribution of the SE tasks were synthesized by \textit{descriptive statistics} and presented in charts. We also used the \textit{narrative synthesis} to summarize the findings in a tabular format and demonstrate both the spectrum of SE tasks and the research depth of specific SE tasks. For \textit{RQ2}, we used \textit{grounded theory} to identify the factors that would have influenced the replicability and reproducibility of ML/DL applications in the SE community. Initially, we identified a set of four candidate factors (\textit{data accessibility}, \textit{source code accessibility}, \textit{data preprocessing}, \textit{hyper-parameter optimization}) based on 100 randomly selected publications that have a direct relation to replicability and reproducibility of ML/DL studies. Through an iterative process of adjusting theory constantly to fit new publications, an additional implicit factor was found (\textit{Understanding Technique Selection} in Section 4.3 B), thus yielding five factors in the end. For \textit{RQ3}, \textit{thematic qualitative} analysis was used to investigate how well authors understand the problem (the SE task to be addressed) as well as their selected ML/DL methods. The ``themes'' ( the level of researchers' understanding of the selected ML/DL techniques and the ingredients to facilitate such understanding) emerged through comparison and categorization. A complementary quantitative synthesis answered this research question with a statistical summary of comparison with different types of data sources for ML/DL models. For \textit{RQ4}, we used \textit{descriptive statistics} to identify the trends of impacts of DL models on SE tasks and 
% as well as their strengths. Then, by comparing the data size of ML and DL applications, we investigated 
the distinctive challenges of applying DL (as opposed to ML) methods, and subsequently suggest future work in this direction.

\section{Results and Synthesis}
\subsection{Overview}
We selected 906 papers\footnote{The full list of included papers will be available upon request.} related to the synergy between ML/DL and SE, with 817 ML and 89 DL studies. For ease of exposition, we will use the term "ML" to refer to those ML studies that involve machine learning but not deep learning and the term "DL" to refer to those ML studies that are deep learning based throughout the paper.

\begin{figure}[t]
  \centering
  \includegraphics[scale = 0.7]{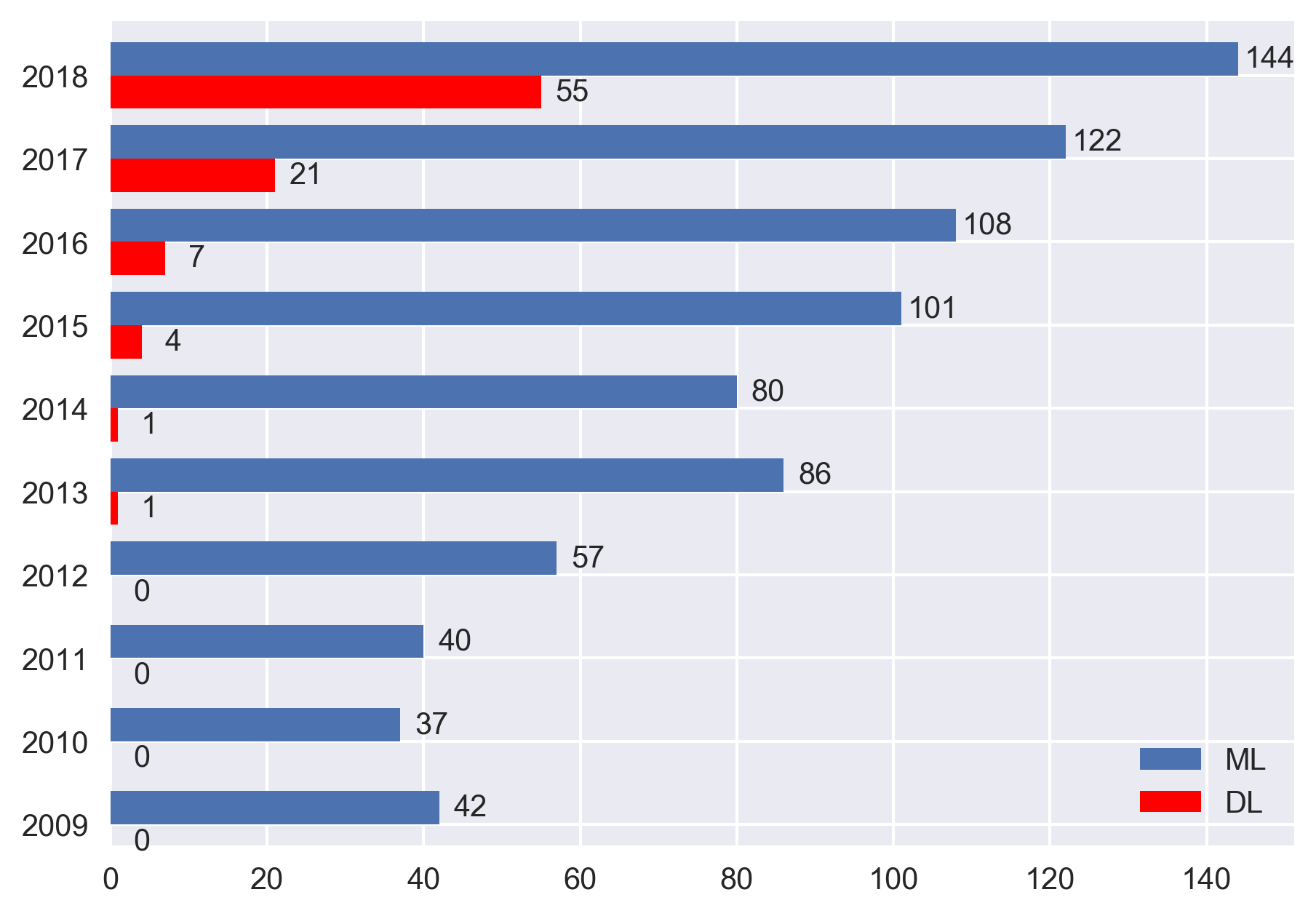}
  \caption{Distribution of papers over year.}
  \label{distribution year}
\end{figure}

Figure \ref{distribution year} shows a significant increase in the number of papers published each year between 2009 and 2018. The blue bar shows the number of ML studies and the red bar shows the number of DL studies. The initial application of DL to SE did not take place until 2013. White et al. \cite{White:2015:TDL:2820518.2820559} introduced DL to software language modeling in 2015, which offered the SE community new ways to learn from source code files to support SE tasks. A possible explanation for this later date is that it took time for SE researchers to digest the emerging DL techniques and cautiously validate their feasibility and effectiveness on SE tasks.

\begin{figure}[t]
  \centering
  \includegraphics[scale = 0.7]{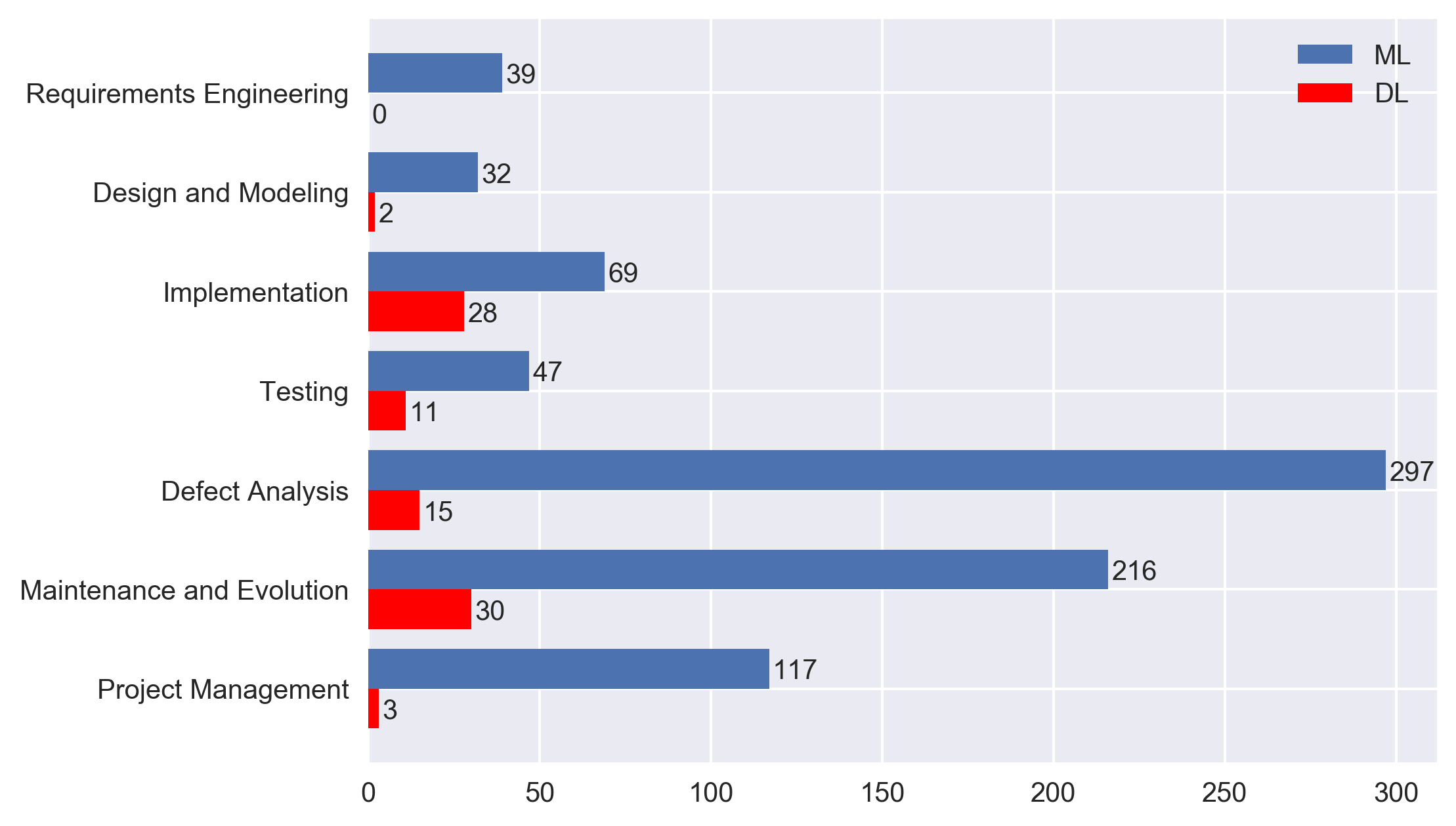}
  \caption{Distribution of papers over SE activities.}
  \label{distribution SE phase}
\end{figure}

Figure \ref{distribution SE phase} presents the distribution of studies over seven SE activities. Among these, \textit{defect analysis} (297+15 ML+DL) and \textit{software maintenance \& evolution} (216+30 ML+DL) take up over half of the collection, mainly because significant amounts of bug reports and software evolution histories are publicly available in the open source repositories. Considering only DL-related studies, \textit{Maintenance \& evolution} is the largest category, while \textit{implementation} contributes the second largest number of DL-related studies (28), % among seven SE activities, 
which seems to suggest that DL models are effective for source code analysis \cite{Hellendoorn:2017:DNN:3106237.3106290}.

\begin{table*}[th!]
  \footnotesize
  \caption{ML/DL for SE: Trend analysis on SE tasks.}
  \label{tab:SE_Summary}
  \begin{tabular}{p{4.5cm} p{7cm} c c c c} 
  %p{1.5cm}  p{1.5cm}}
    
    \hline
     \multirow{2}{2 cm}{\textbf{SE Activity}} & \multirow{2}{2 cm}{\textbf{SE Task}} & \multicolumn{2}{c}{\textbf{2009 -2014}} & \multicolumn{2}{c}{\textbf{2015 -2018}}\\\cline{3-6}
    
    \centering & & \textbf{ML} & \textbf{DL} & \textbf{ML} & \textbf{DL}\\
     \hline
   
    Requirements Engineering (39) 
    & \textcolor{blue}{R1.Requirements Tracing (4)}& 2 & 0 & 2 & 0\\
    & \textcolor{blue}{R2.Requirements Detection and Classification (29)}& 15 & 0 & 14 & 0\\
    & \textcolor{blue}{R3.Requirements Prioritization (5) }& 3 & 0 & 2 & 0\\ 
    & \textcolor{red}{R4.User Story Detection (1) }& 0 & 0 & 1 & 0\\ \hline
    %{\color{blue} R1(2), R2(15), R3(3)} 
    %& {\color{blue} R1(2), R2(14), R3(2)}, \newline {\color{red}R4(1)}\\
    
    Design and Modeling (34) 
    & \textcolor{red}{D1.Architecture Tactics Detection (4)} & 0 & 0 & 4 & 0\\ 
    & \textcolor{red}{D2.Data and Service Modeling (8)} & 0 & 0 & 7 & 1 \\ 
    & \textcolor{blue}{D3.Model Optimization (9)} & 6 & 0 & 2 & 1 \\
    & \textcolor{blue}{D4.Process Model Recommendation (2)} & 1 & 0 & 1 & 0\\
    & \textcolor{blue}{D5.Agent Behavior Modeling (7)} & 4 & 0 & 3 & 0\\
    & \textcolor{blue}{D6.Design Pattern Detection (4)} & 2 & 0 & 2 & 0\\ \hline 
    %& {\color{blue} D3(7), D6(2), D5(4), D4(1)} 
    %& {\color{blue} D3(3), D4(1), D5(3), D6(2)}, {\color{red} D1(4), D2(8)}\\
    
    Implementation (97) 
    & \textcolor{blue}{I1.Code Optimization (20)} & 3 & 0 & 9 & 8 \\
    & \textcolor{red}{I2.Code Summarization (7)} & 0 & 0 & 0 & 7 \\
    & \textcolor{blue}{I3.API Mining (16)} & 3 & 0 & 12 & 1 \\
    & \textcolor{blue}{I4.Code Generation and Completion (16)} & 5 & 0 & 4 & 7 \\
    & \textcolor{red}{I5.Source Code Identification (2)} & 0 & 0 & 0 & 2 \\
    & \textcolor{blue}{I6.Program Synthesis (8)} & 2 & 0 & 5 & 1 \\
    & \textcolor{blue}{I7.Code Smell/Anti-pattern Detection (17)} & 5 & 0 & 10 & 2 \\ 
    & \textcolor{blue}{I8.Software Modularization (4)} & 2 & 0 & 2 & 0\\
    & \textcolor{blue}{I9.Code Comments Classification (7)} & 1 & 0 & 6 & 0\\ \hline
    %\newline {\color{blue} I1(3), I3(3), I4(5), \newline I6(2), I7(5), I8(2), \newline I9(1)} 
    %& {\color{blue} I1(17), I3(13), I4(11), \newline I6(6), I7(12), I8(2), \newline I9(6)}, {\color{red} I2(7), I5(2)}\\
    
    Testing (48) 
    & \textcolor{blue}{T1.Test Case Generation (14)} & 5 & 0 & 6 & 3 \\
    & \textcolor{blue}{T2.Test Automation (17)} & 8 & 0 & 8 & 1 \\ 
    & \textcolor{red}{T3.Test Report Classification (3)} & 0 & 0 & 3 & 0\\
    & \textcolor{blue}{T4.Test Prioritization and Optimization (14)} & 5 & 0 & 9 & 0\\ \hline
    %& {\color{blue} T1(6), T2(8), T4(5)} 
    %& {\color{blue} T1(9), T2(9), T4(9)}, \newline {\color{red} T3(3)}\\
    
    Defect Analysis (309) 
    & \textcolor{blue}{A1.Defect Prediction (187)} & 75 & 0 & 105 & 7 \\
    & \textcolor{blue}{A2.Defect Detection and Localization (68)} & 21 & 0 & 40 & 7\\
    & \textcolor{blue}{A3.Defect Categorization (12)} & 7 & 0 & 5 & 0 \\
    & \textcolor{blue}{A4.Bug Report Classification (15)} & 5 & 0 & 10 & 0\\
    & \textcolor{blue}{A5.Bug Assignment (21)} & 14 & 0 & 7 & 0\\
    & \textcolor{blue}{A6.Duplicate Bug Report Detection (5)} & 2 & 0 & 3 & 0\\
    & A7.Defect Escalation Prediction (1) & 1 & 0 & 0 & 0\\ \hline
    %& {\color{blue} A1(75), A2(21), A3(8), \newline A4(5), A5(14), A6(2)}, \newline A7(1)
    %& {\color{blue} A1(112), A2(48), A3(7), \newline A4(10), A5(7), A6(3)}\\
    
    Maintenance and Evolution (244) 
    & \textcolor{blue}{M1.Repository Mining (49)} & 13 & 0 & 30 & 6 \\
    & \textcolor{blue}{M2.Sentiment Analysis (26)} & 3 & 1 & 19 & 3 \\
    & \textcolor{blue}{M3.Code Clone and Similarity Detection (16)} & 4 & 0 & 5 & 7 \\ 
    & M4.Authorship Attribution (1) & 1 & 0 & 0 & 0\\
    & \textcolor{blue}{M5.Software Change Prediction (33)} & 10 & 0 & 23 & 0 \\
    & \textcolor{blue}{M6.Defect Fixing (13)} & 6 & 0 & 5 & 2 \\
    & \textcolor{blue}{M7.Software Application/Artifact Categorization (14)} & 7 & 0 & 5 & 2\\
    & \textcolor{red}{M8.Software Refactoring (2)} & 0 & 0 & 2 & 0\\
    & \textcolor{blue}{M9.Software Quality Prediction (40)} & 18 & 0 & 19 & 3\\
    & \textcolor{red}{M10.Specification Mining (6)} & 0 & 0 & 4 & 2\\ 
    & M11.Aspect Mining (1) & 1 & 0 & 0 & 0\\
    & \textcolor{blue}{M12.Configuration Optimization (6)} & 4 & 0 & 2 & 0\\
    & \textcolor{blue}{M13.Program Comprehension (9)} & 2 & 0 & 7 & 0\\
    & \textcolor{blue}{M14.Tag Recommendation (2)} & 1 & 0 & 1 & 0\\
    & \textcolor{blue}{M15.Traceability link generation and recovering (12)} & 4 & 0 & 7 & 1\\
    & \textcolor{blue}{M16.Report/Review Summarization (6)} & 3 & 0 & 2 & 1\\ 
    & \textcolor{red}{M17.Reputation Management (1)} & 0 & 0 & 1 & 0\\
    & \textcolor{red}{M18.Developers' Behavior Analysis (6)} & 0 & 0 & 6 & 0\\ 
    & M19.Copyright Authentication (1) & 1 & 0 & 0 & 0\\ \hline
    %& {\color{blue} M1(13), M2(4), M3(4), \newline M5(10), M6(6), M7(7), \newline M9(18), M12(4), M13(2), \newline M14(1), M15(4), M16(3)}, \newline M4(1), M11(1), M19(1) 
    %& {\color{blue} M1(36), M2(22), M3(12), \newline M5(23), M6(7), M7(7), \newline M9(23), M12(2), M13(7), \newline M14(1), M15(8), M16(2)}, \newline {\color{red} M8(2), M10(6), M17(1), \newline M18(6)}\\
    
    Project Management (120) 
    & \textcolor{blue}{P1.Schedule/Effort/Cost Estimation (74)} & 41 & 0 & 32 & 1\\
    & \textcolor{red}{P2.Software Size Estimation (1)} & 0 & 0 & 1 & 0 \\ 
    & \textcolor{blue}{P3.Process Management (8)} & 2 & 0 & 6 & 0\\
    & \textcolor{blue}{P4.Energy Estimation (5)} & 1 & 0 & 3 & 1\\
    & \textcolor{red}{P5.Risk Prediction (2)} & 0 & 0 & 2 & 0\\
    & \textcolor{blue}{P6.Performance Prediction (19)} & 11 & 0 & 7 & 1\\
    & P7.Task Dependency Detection (1) & 1 & 0 & 0 & 0\\
    & \textcolor{blue}{P8.Project Outcome Prediction (4)} & 2 & 0 & 2 & 0\\
    & \textcolor{blue}{P9.Data Protection (6)} & 5 & 0 & 1 & 0\\
    %& {\color{blue} P1(41), P3(2), P4(1), \newline P6(11), P8(2), p9(5)}, \newline p7(1) 
    %& {\color{blue} P1(33), P3(6), P4(4), \newline P6(8), P8(2), p9(1)}, \newline {\color{red} P2(1), P5(2)}\\
    
    \hline
\end{tabular}
\end{table*}

\begin{table*}[t]
\footnotesize
  \renewcommand{\arraystretch}{2}
  \caption{SE for ML/DL: Summary.}
  \label{tab:SE_for_ML}
  \begin{center}
  \begin{tabular}{p{2.5cm} p{3cm} p{3cm} p{3cm} p{2cm}}
    \hline
    \textbf{SE Activity} & \textbf{SE Task} & \textbf{SE Technique} & \textbf{ML/DL Workflow} & References\\
    \hline
        Testing (10) & T5.DL/ML model testing (9) \newline T1.Test Case Generation(1) & Metamorphic/Mutation Testing \newline Test Selection & Model Evaluation & \cite{XIE2011544,Tian:2018:DAT:3180155.3180220,Dwarakanath:2018:IIB:3213846.3213858,Ma18,Groce14,Ma:2018:DMT:3238147.3238202,Zhang:2018:DGM:3238147.3238187,Cheng18,Sun:2018:CTD:3238147.3238172,XieHo09}\\
        Defect Analysis (3) & A3.Defect Categorization (3) & Manual Analysis & Model Evaluation & \cite{Zhang:2018:EST:3213846.3213866,Sun17,Thung12}\\
        Maintenance \newline and Evolution (2) & M20.DL System Integration and Deployment (2) & OSGi/Scaling & Deployment & \cite{CONINCK201852,Coates:2013:DLC:3042817.3043086}\\
  \hline
\end{tabular}
\end{center}
\end{table*}

\subsection{Trend Analysis for SE and ML/DL (RQ1)}
We analyzed the impact that ML/DL and SE have on each other: \textit{ML/DL for SE} and \textit{SE for ML/DL}. In addition, based on how ML/DL-based techniques generate impacts on diverse SE tasks and how SE methods contribute to ML/DL-based systems, we categorized the 906 studies into a total of 60 distinct SE tasks over the aforementioned seven SE activities. 891 (98\%) of the 906 studies concern \textit{ML/DL for SE}, covering 58 (97\%) of the 60 distinct SE tasks. The remaining 15 studies concern \textit{SE for ML/DL}, covering five of the 60 (8\%) distinct SE tasks.

\subsubsection{ML/DL for SE}
To investigate both the \textit{breadth} and the \textit{depth} of the changes that emerging ML/DL techniques have brought to SE, we set 2015 as the line of demarcation to compare studies published in two time periods, 2009-2014 and 2015-2018, as 2015 is the year DL studies became prevalent in SE. In Table \ref{tab:SE_Summary}, the columns ``2009-2014'' and ``2015-2018'' list the number of ML/DL-based studies that were published within each of the two time periods for different SE tasks. The total number of relevant studies is shown in parentheses beside each SE task. The SE tasks are color coded in blue, red and black. 
%Blue shows the SE tasks that 
The blue tasks were continuously studied and experimented with ML/DL methods in both time periods (2009-2014 and 2015-2018). 
%Red shows the newly supported SE tasks 
The red tasks are those newly supported by ML/DL from 2015 to 2018. 
%Black indicates that 
The black tasks are those where the trials of ML/DL techniques on these SE tasks only occurred in 2009-2014.

\textit{\textbf{What impacts do emerging ML/DL bring to SE?}} In general, we observed a rapid growth of the number of SE tasks that can be supported and automated by ML/DL techniques in the past decade. Based on the 12 SE tasks that were newly experimented with emerging ML/DL methods after 2014 (the red tasks in Table \ref{tab:SE_Summary}), we discovered two impacts that emerging ML/DL techniques bring to SE (\textit{breadth}).
First, \textbf{a larger variety of SE artifacts has been effectively analyzed using ML/DL techniques to improve the productivity of the development processes}, including \textit{user stories (R4)}, \textit{architecture tactics (D1)}, \textit{test reports (T3)}, \textit{formal specifications (M10)} and even \textit{developers' interaction data (M18)}. For example, Bao et al. \cite{Bao2018} used a Condition Random Field (CRF)-based ML approach to infer a set of basic development activities in real world settings by analyzing developers' low-level actions.
Second, as shown in Table \ref{tab:SE_Summary}, \textbf{two SE tasks (\textit{Code Summarization (I2)} and \textit{Source Code Identification (I5)}) that were not successfully tackled by traditional ML methods have been addressed by emerging DL techniques 
%in the SE community 
since 2015}.
For \textit{Source Code Identification (I5)}, through the application of CNNs, SE researchers have made significant progress on identifying and locating the source code embedded in multimedia (image and video) artifacts. For example, Ott et al. \cite{Ott18} developed a DL solution based on CNNs to classify the presence or absence of code in video tutorials. Alahmadi et al. \cite{Alahmadi:2018:APL:3273934.3273935} were able to accurately locate the source code in video frames by CNNs as well. The contributions in both studies set important steps towards the ultimate extraction of relevant code from videos, which will expand SE data sources and give developers access to a richer set of documented source code that is currently not leveraged.
For \textit{Code Summarization (I2)}, employing the  DL-based Encoder-Decoder framework (see Section 4.5.2 for details), SE researchers are able to overcome the performance bottleneck of traditional SE methods, resulting in automatically generating natural language descriptions such as code comments and commit messages. For instance, Hu et al. \cite{Hu:2018:DCC:3196321.3196334} proposed an Attentional RNN Encoder-Decoder model to automatically generate code comments for Java methods, which overcame the limitations of previous approaches that involve manually crafted templates or information retrieval, such as the failure to capture the semantic relationship between source code and natural language description.

We also witnessed %that 
the ability of emerging ML/DL techniques %offer new capabilities to have enabled SE researchers and practitioners 
to overcome the challenges that cannot be addressed with traditional SE practices.
Throughout the last decade, 41 SE tasks (shown in blue in Table \ref{tab:SE_Summary}) spanned both time periods, which implies the continuous and long-term use of ML/DL to 
%explore additional capabilities and 
acquire deeper insights into the same SE tasks (\textit{depth}). In particular, \textbf{we observed an evolution from ML- to DL-based solutions for 23 SE tasks, mainly due to the replacement of the hand-crafted features required by ML with representation learning by DL}. Two examples of SE tasks are illustrated below to show this trend. Contributing the largest number of papers (187), \textit{defect prediction (A1)} has been the most favorable research task %under research 
in the SE community, motivated by the potential cost and quality rewards associated with accurately predicting defect-prone software \cite{Shepperd14}. Before 2015, traditional ML classifiers were the dominant solutions to software defect prediction (75 papers) \cite{Ghotra:2015:RIC:2818754.2818850}. These classifiers were mainly trained on hand-crafted features, which sometimes failed to capture the semantics of programs. To address this problem, DL-based approaches have been widely adopted (7 papers) to generate more expressive, complicated and nonlinear features from the initial feature sets \cite{Yang15} or directly from source code \cite{Wang16, Li17}.
% which are more capable of expressing the nature of a problem. 
\textit{Code Clone and Similarity Detection (M3)} is another common SE task whose initial capability was limited to %being able to 
detecting only Type I-III clones based on the textual similarity computed from hand-crafted features. It was then augmented to spotting all four types of clones using both textual and functional similarities through source code representation learned via DL \cite{White:2016:DLC:2970276.2970326}. 

\subsubsection{SE for ML/DL}
\textit{\textbf{How do SE practices contribute to developing and evolving ML/DL systems?}} The study by Microsoft \cite{Amershi:2019:SEM:3339914.3339967} discovered three fundamental differences between the AI and SE domains, which imply the need to adjust SE practices when they are applied to ML/DL system development and evolution. The study also introduced the nine stages of the ML workflow \cite{Amershi:2019:SEM:3339914.3339967}. In Table \ref{tab:SE_for_ML}, we identified 15 studies that attempted to integrate traditional SE techniques into two stages of the ML/DL workflow, Model Evaluation (T1, T5, A3) and Deployment (M20). Among them, nine (60\%) of the 15 studies proposed their testing criteria for measuring the quality of the test data and identified the implementation bugs of ML/DL systems using various SE testing techniques, including metamorphic testing, mutation testing and concolic testing. In addition, one study \cite{Groce14} proposed and formally defined three test selection methods for end users of interactive ML systems. We also identified that three (20\%) of the 15 studies investigated the characteristics of ML/DL defects, examined the root cause \cite{Zhang:2018:EST:3213846.3213866} and analyzed the relations with defect fixing \cite{Sun17,Thung12}. Apart from \textit{Testing} and \textit{Defect Analysis}, two more studies focused on scaling DL %systems 
to very large models and training sets during the \textit{Maintenance and Evolution} phase. One study \cite{CONINCK201852} implemented a modular framework to tackle the lack of distributed training for most DL frameworks, with a clear focus on dynamic distributed training and deployment of deep neural networks on a multitude of heterogeneous devices. Another study \cite{Coates:2013:DLC:3042817.3043086} presented details of a very large scale DL system on high performance computing infrastructure.

\textit{\textbf{What are the challenges involved in applying software testing and debugging practices to ML/DL systems, and what caused such challenges?}} 13 (87\%) of the 15 studies applied SE methods to test and debug (\textit{Testing} + \textit{Defect Analysis}) ML/DL systems. Compared to traditional software systems, ML/DL systems are relatively less deterministic and more statistics-orientated \cite{zhang2019machine} due to their fundamentally different nature and construction. Admittedly, testing and debugging ML/DL systems is still at an early stage, mainly relying on probing the accuracy on test data that are randomly drawn from manually labeled data and ad hoc simulations \cite{Ma18}. We observed five causes that may lead to the challenges involved in developing a systematic, effective and efficient testing framework for ML/DL systems. 
First, due to the large number of inputs and the uncertainties of the ML/DL model context to cover all the operational scenarios, test cases can be too complex and time consuming to create \cite{Dwarakanath:2018:IIB:3213846.3213858}. 
Second, because of the stochastic nature of the learning process, it is hard to determine the test oracle \cite{Dwarakanath:2018:IIB:3213846.3213858}. 
Third, because there can be multiple underlying reasons, it is difficult to confirm a bug \cite{Dwarakanath:2018:IIB:3213846.3213858}.
Fourth, ML/DL programs tend to be more resistant to computational mistakes, so coincidental correctness\footnote{Coincidental correctness refers to the following phenomenon: though bugs are triggered during the training process, the trained model still achieves the desirable accuracy on the test set. The bugs were finally discovered by code review.} may occur \cite{Zhang:2018:EST:3213846.3213866}.
Fifth, there is still a lack of systematic evaluation of how different assessment metrics are correlated with each other, and how these assessment metrics are correlated with the fault-revealing ability of tests \cite{zhang2019machine}. 

\textit{\textbf{What are the potential research opportunities of SE practices for ML/DL systems?}} Among the 15 collected papers, we observed three research trends of SE practices that are applied to ML/DL systems: (1) the aforementioned statistics reveal that ML/DL testing and debugging practices are the dominant applications in SE; (2) following the surge of DL as shown in Figure \ref{distribution year}, the application of SE practices is gradually shifting from ML to DL systems: 10 (67\%) of the 15 papers present techniques that are specifically designed for DL systems and the remaining five papers concern the development of general ML systems; and (3) when classifying the papers based on the four learning paradigms mentioned in Section 2, all 15 papers focused on supervised learning systems. In light of these trends, many research opportunities remain:
\begin{itemize}
    \item \textit{Expanding application scenarios}: In addition to devoting effort to addressing challenges in Model Evaluation (testing and debugging ML/DL systems), SE practices can be applied to different stages of the ML/DL workflow. For example, in the model requirements stage \cite{Amershi:2019:SEM:3339914.3339967}, requirement traceability activities would allow us to verify whether features are feasible to implement with ML/DL, which features are useful for an existing/new product, and what types of models are most appropriate for the given problem.
    \item \textit{Applying SE practices to different kinds of ML/DL sytems}:
    As mentioned above, all the 15 papers used in our study involve %can be classified as belonging to 
    the supervised learning paradigm. Future research efforts can examine how SE practices can be applied to the development of UL or RL-based systems.
    %For all the 906 papers that we have identified, the majority of papers primarily focused on supervised learning systems; however, we still observed that 87 (10\%) of the 906 papers utilized unsupervised learning (UL), reinforcement learning (RL) or hybrid (i.e., contain both) approaches in many SE tasks across the seven SE activities. Therefore, more research efforts on UL and RL systems would be beneficial.
    
\end{itemize}

\begin{tcolorbox}[
    title = {Summary of RQ1}
    ]
We identified two impacts that emerging ML/DL techniques have brought to SE and observed an evolution from ML- to DL-based solutions for 23 SE tasks. At the same time, we identified three research trends of SE practices applied to ML/DL systems, and five causes that may lead to the challenges involved in developing a systematic testing framework for ML/DL systems.
\end{tcolorbox}

\begin{figure}[t]
  \centering
  \includegraphics[scale = 0.9]{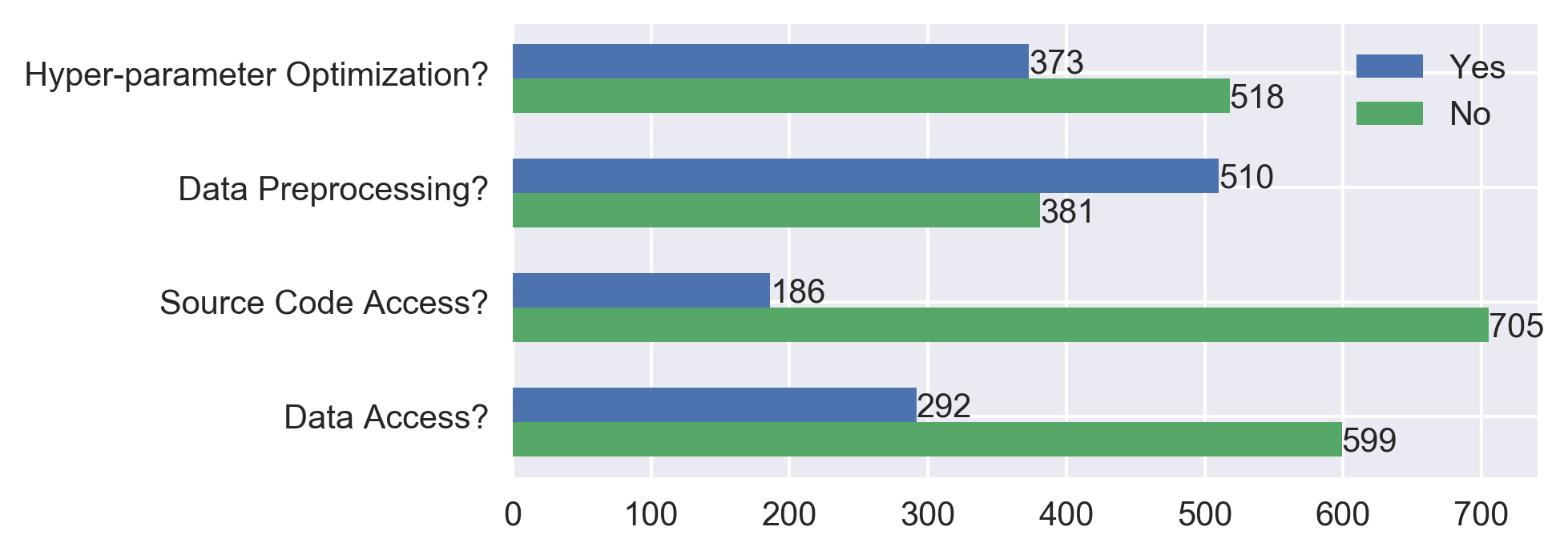}
  \caption{Factors of replicability and reproducibility.}
  \label{factor}
\end{figure}

\subsection{Factors that Influence Replicability and Reproducibility (RQ2)}

\textit{\textbf{Why are replicability and reproducibility of studies essential to SE?}}
The replicability and reproducibility of ML/DL applications have a great impact on the transfer of research results into industry practices. According to the ACM policy on artifact review and badging \cite{result2017artifact}, replicability refers to the ability of an independent group to obtain the same result using the author's own artifacts. Likewise, reproducibility is the act of obtaining the same result without the use of original artifacts (but generally using the same methods). Reproducibility is clearly the ultimate goal, but replicability is an intermediate step to promote practices. Somewhat unfortunately, according to Fu and Menzies \cite{Fu:2017:EOH:3106237.3106256}, it is hard to replicate or reproduce ML/DL applications from SE research due to the nondisclosure of datasets and source code. 

\textit{\textbf{How many studies addressing the synergies between ML/DL and SE were replicated and reproduced?}} We intentionally excluded studies published in 2018 and found that only 90 (13\%) of the 707 studies (2009-2017) in our collection were either replicated as baselines or reproduced for other datasets. We then grouped all 906 studies into three classes based on how they were evaluated. Among the 906 studies, 805 (89\%) were evaluated using open source datasets, 61 (7\%) using industrial datasets and 40 (4\%) using student collections\footnote{A student collection is a dataset from an exclusive source, such as a student submission, a survey or a field study.}. Although the vast majority of evaluations on publicly available open source datasets imply an awareness of the importance of replicability in the SE community, only 90 (13\%) of the 707 studies (2009-2017) were replicated or reproduced. On the other hand, the reproducibility from academic to industrial settings remains a significant challenge, with only 7\% of the studies being transferred to industrial practices. 

\textit{\textbf{What are the factors that influence the replicability and reproducibility of ML/DL-related studies in SE?}} To answer this question, we included 891 of the 906 papers for further analysis, specifically excluding the 15 studies that involve \textit{SE for ML/DL}. Following a \textit{grounded theory} approach in an iterative manner, we identified a set of five factors that deserve SE researchers' attention as far as improving the replicability and reproducibility of ML/DL-related SE studies is concerned. We considered two kinds of artifacts, data and source code, whose accessibility is crucial to replicability. Reproducibility requires not only the availability and accessibility of artifacts but also an understanding of the original studies and the appropriate management of data preparation and configuration of model parameters.

\vspace{2.5mm}\noindent\textbf{A. Factors for Replicability}\vspace{2.5mm}

\textit{Data Accessibility.}
Data comprises the training and test data used by ML/DL algorithms after preprocessing. Only 292 (33\%) of the 891 studies provided DOI or links to their data\footnote{According to the ACM policy, an artifact is available if a DOI or a link to the data or source code repository, along with a unique identifier for the object, is provided.}, as shown in Figure \ref{factor}. If the data is not accessible, there may be a mismatch between the regenerated data and the originally preprocessed data. This mismatch may make it difficult to replicate the experimental results. Specifically, it would be difficult to locate the problem when the replicated results are far from the results reported in the original study.

\textit{Source Code Accessibility.}
The full package of source code consists of at least two parts, end-to-end scripting (e.g., data preprocessing, statistical analyses) and the implementation of the ML/DL-based approach (e.g., model construction, parameter tuning methods, training algorithms) \cite{broman2017recommendations}, which guarantee the compliance with the same ML/DL workflow when replicating. Figure \ref{factor} shows that 705 (79\% ) of the 891 studies did not share any DOI or links to the package of source code, which leaves follow-up studies no choice but to re-implement the entire approach from scratch by relying on the description of the approach in the original study. The tremendous effort involved in
replicating previous studies could hamper the progress of research evolution. %According to another statistic, 
Moreover, as many as 561 (63\%) of the 891 studies did not publish either the preprocessed data or the package of source code, causing a potential dilemma for replicability.

\begin{figure}[t]
  \centering
  \includegraphics[scale = 0.7]{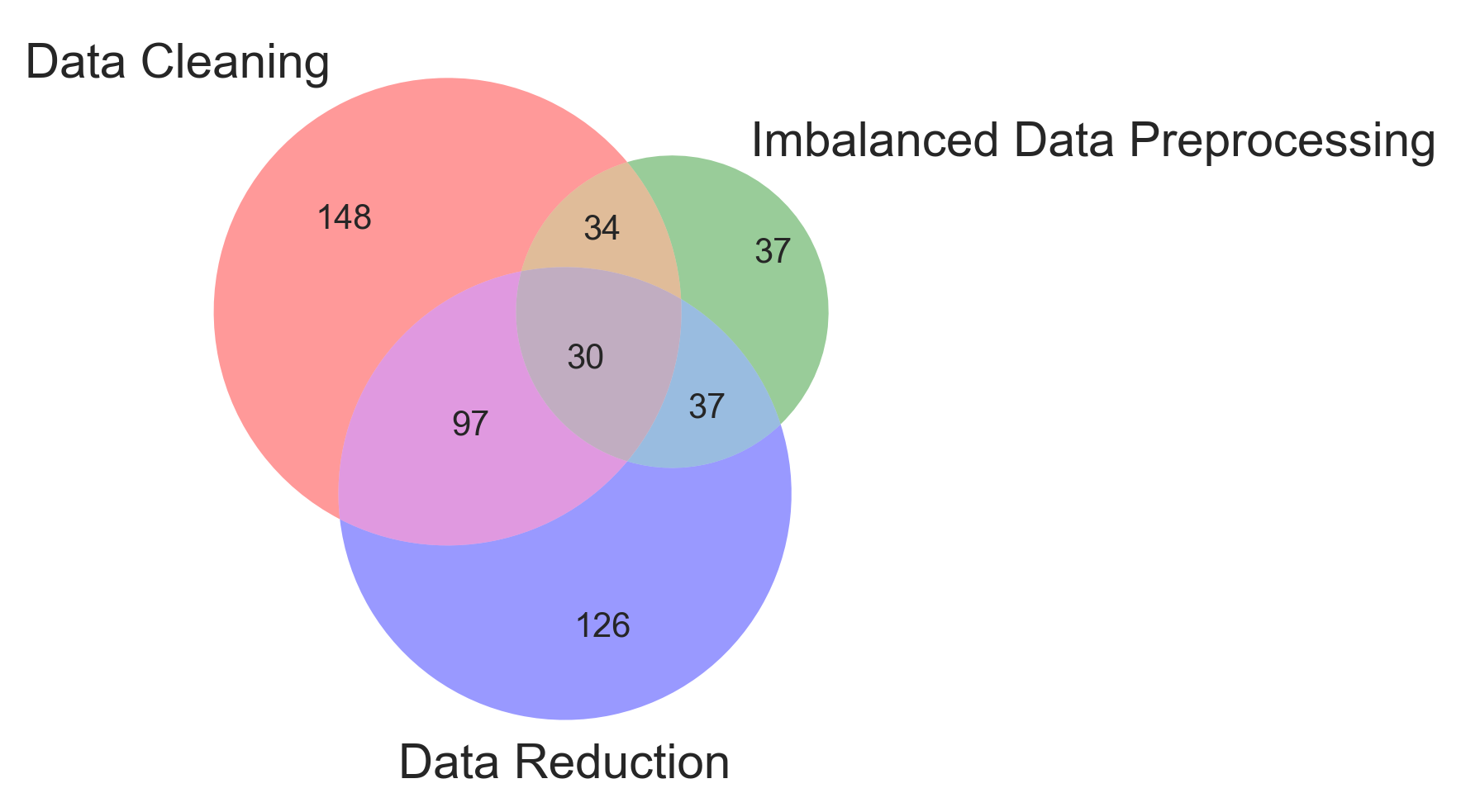}
  \caption{Data preprocessing.}
  \label{data_preprocess}
\end{figure}

\begin{figure}[t]
  \centering
  \includegraphics[scale = 0.7]{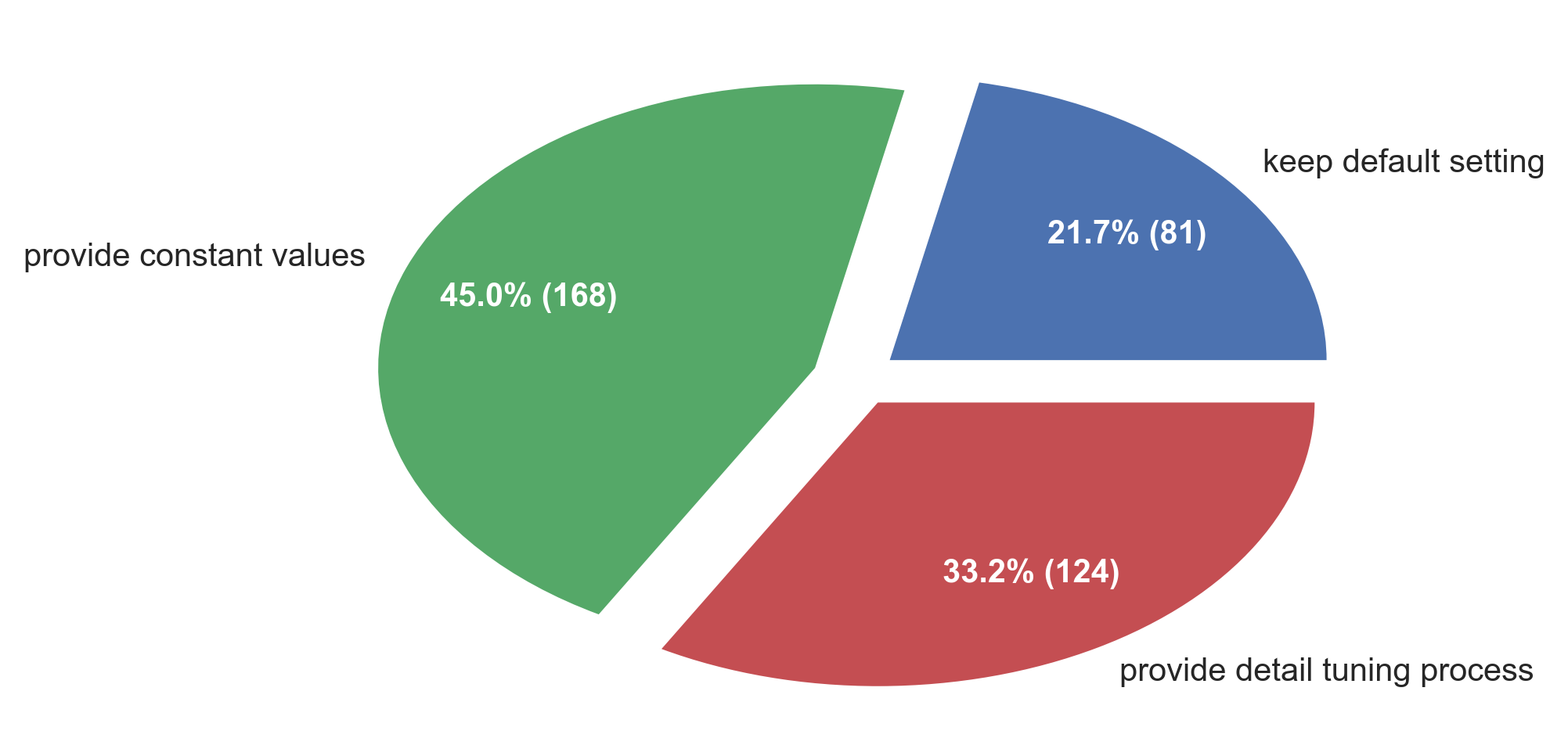}
  \caption{Hyper-parameter optimization.}
  \label{Hyper-parameter}
\end{figure}

\vspace{2.5mm}\noindent\textbf{B. Factors for Reproducibility}\vspace{2.5mm}

\textit{Data Preprocessing.}
There are three types of data preprocessing techniques for ML/DL: data cleaning (noise filtering, missing value imputation), imbalanced data preprocessing, and data reduction (e.g., feature selection, space transformation) \cite{GARCIA20161}. Data preprocessing is indispensable to ML/DL because a dataset is likely to contain various problems, such as inconsistencies, errors, out-of-range values, impossible data combinations, and missing values, making it unsuitable to start a ML/DL workflow \cite{GARCIA20161}. Alternatively, datasets (especially industrial proprietary data and open source data) are usually different from each other, thus calling for extra caution when selecting appropriate types of data preprocessing methods that match the datasets. More than half of the studies (510) used at least one type of data preprocessing method shown in Figure \ref{factor}. The Venn Diagram in Figure \ref{data_preprocess} shows the distribution of data preprocessing methods used in the 510 studies. We noticed two potential threats to
%that can potentially affect 
reproducibility. First, as shown in Figure \ref{data_preprocess}, researchers tend to overlook the preprocessing of imbalanced data, which may downgrade system performance when reproducing a study on a new dataset. For example, the accuracy of defect prediction models is often adversely affected by the imbalanced nature of the software defect datasets, in which there are naturally fewer instances of the defective classes than those of the non-defective classes \cite{Bennin18}. Second, during data cleaning, 309 (61\%) of the 510 studies (148+34+30+97 with Data Cleaning in Figure \ref{data_preprocess}) only mentioned the removal of stop words, with no explicit report of what further actions (e.g., outlier filtering) were taken and to what extent. Furthermore, although most studies reported the raw data size, none of the 309 studies reported the data size difference before and after cleaning.
%potentially misleading others when applying to an entirely different dataset.

\textit{Hyper-parameter Optimization} refers to the process of choosing a set of optimal hyper-parameters for a ML/DL algorithm. Although parameter tuning is typically regarded as a ``black art'', its impact is well understood \cite{bergstra2012random} and tuning needs to be repeated whenever the data or the goals are changed \cite{FU2016135}. Figure \ref{factor} shows that 518 (58\%) of the 891 studies chose not to even mention hyper-parameters. As shown in Figure \ref{Hyper-parameter}, among the remaining 373 studies, only 124 (33\%) either used existing optimization techniques (e.g., Grid Search) or described manual search strategies (a range of values for each hyper-parameter to be tuned), which are useful for reproducing their research results. Merely using the default parameter setting is not recommended \cite{FU2016135}, the reason being that hyper-parameter optimization often plays a key role in system performance and the optimal parameter setting for different tasks can be very different. 
%which means constant values of hyper-parameters are useless as well.

\begin{figure}[t]
  \centering
  \includegraphics[scale = 0.8]{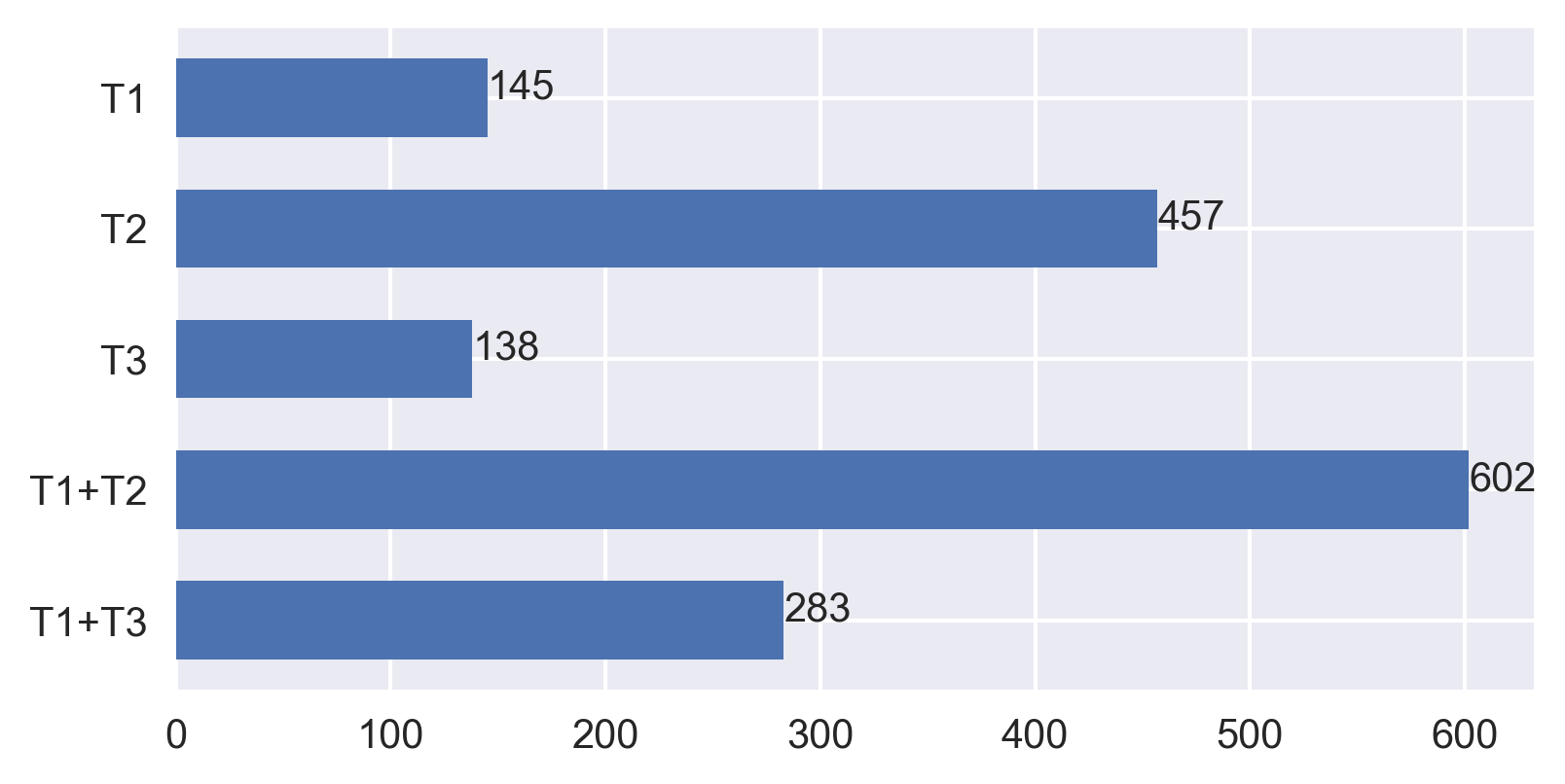}
  \caption{Distribution of different types of rationales.}
  \label{rationale}
\end{figure}

\textit{Understanding Technique Selection.}
Understanding why a ML/DL technique was selected to solve a specific SE problem would be helpful for reproducing the study in more generalized scenarios. Section 4.4 will elaborate on this factor.

\begin{tcolorbox}[
    title = {Summary of RQ2}
    ]
We identified two factors (\textit{Data Accessibility}, \textit{Source code Accessibility}) that influenced replicability and three factors (\textit{Data Preprocessing}, \textit{Hyper-parameter Optimization}, \textit{Understanding Technique Selection}) that influenced the reproducibility of ML/DL-based studies in SE.
\end{tcolorbox}

\subsection{Understanding ML/DL Technique Selection (RQ3)}

It is important to ensure that others understand why a ML/DL technique is selected for a specific SE problem because it would improve the generalizability and applicability of the approach. \textit{RQ3} concerns the question of what ingredients in a study would facilitate the understanding of why a ML/DL algorithm was selected to tackle a specific SE problem. To address \textit{RQ3}, we used the \textit{thematic synthesis} method to identify patterns from 891 of the 906 papers for further analysis, specifically excluding the 15 studies on \textit{SE for ML/DL}. We then classified the extracted rationale from identified studies into three types, \textit{T1}, \textit{T2}, \textit{T3}. 

\begin{itemize}
    \item \textit{T1}: The studies only justified the algorithm selection based on the performance of the algorithm from historical studies or empirical reports. For example, a typical statement could be ``\textit{We selected random forest since this algorithm has been used in many previous empirical studies and tends to have good predictive power}''. 
    \item \textit{T2}: Though not providing any explicit rationales, some studies designed experiments for technique comparisons that demonstrated that the selected algorithms performed better than other methods. For instance, Liu et al. \cite{10.1145/3238147.3238216} conducted an empirical evaluation to compare \textit{RecRank}, a discriminative SVM-based API recommendation approach that they proposed, with two baseline systems, \textit{Gralan} \cite{7194632} (a statistical generative model for API recommendation) and \textit{APIREC}, a state-of-the-art API recommendation approach that was proposed by \textit{Gralan}'s authors \cite{10.1145/2950290.2950333}. The comparison results showed that \textit{RecRank} significantly improves %top-1
    API recommendation accuracy.
    %and mean reciprocal rank.
    \item \textit{T3}: The studies justified the selected algorithms based on their ability to overcome the obstacles associated with a specific SE task. For example, Han et al. \cite{Han17} applied DL to predict the severity of software vulnerability because of its ``\textit{ability to automatically learn word and sentence representations for the prediction task, thus removing the need for manual feature engineering which is a challenging empirical process due to the diversity of software vulnerability and the richness of information in vulnerability description}''.
\end{itemize}
 
\textit{\textbf{How many studies provided the three rationales (T1, T2, and T3)?}} As shown in Figure \ref{rationale}, we noticed that 283 (\textit{T1+T3}, 32\%) of the 891 studies explicitly justified the rationales behind the selection of a ML/DL algorithm to various degrees. Moreover, 602 (\textit{T1+T2}, 68\%) of the 891 studies provided rationales belonging to \textit{T1} or \textit{T2}, which implies that many SE studies considered performance as the top criterion for selecting ML/DL algorithms. This potentially has two negative impacts on the SE community. First, researchers may blindly trust the good performance of specific ML/DL techniques while ignoring their downsides, such as the long computation time and higher modeling complexity, thus missing a trade-off analysis among techniques \cite{Jiang:2017:AGC:3155562.3155583, Fu:2017:EOH:3106237.3106256, Menzies18, Xu:2018:PRS:3239235.3240503}. Second, failure to provide a detailed explanation of the applied ML/DL technique(s) in a study may make it hard for others to determine the suitability of employing the method for a different problem. On the other hand, 138 studies provided rationales belonging to \textit{T3} (Figure \ref{rationale}). However, we also noted that a majority of these 138 studies only provided a general description of the advantages of the selected algorithms without any explanation of how to implement the approach. Hence, it remains difficult for others to make changes to the original implementation to suit different needs during the reproduction process. An ideal example, successfully being replicated or reproduced three times, is that of Wang et al. \cite{Wang16}, who provided the rationales when determining which parts of source code needed to be parsed, which mislabeling data detection approach was appropriate and which DL algorithm was selected. This finding leads to another interesting question:

\textit{\textbf{What ingredients should be taken into consideration in order to better understand the selected ML/DL techniques for a specific SE task?}}  The generalization of ML/DL-based approaches is a major concern for many studies. Since every technique has its own limitations in different deployment contexts, it would be useful to discuss what are the pitfalls and how to avoid and/or mitigate the issues in ML/DL experimental design. We found that three ingredients in the description of a methodology were quite helpful for improving the generalizability of a technique: (1) feature preparation, (2) the underlying assumptions, and (3) the error analysis conducted as part of model evaluation. 

\begin{figure}[t]
  \centering
  \includegraphics[scale = 0.8]{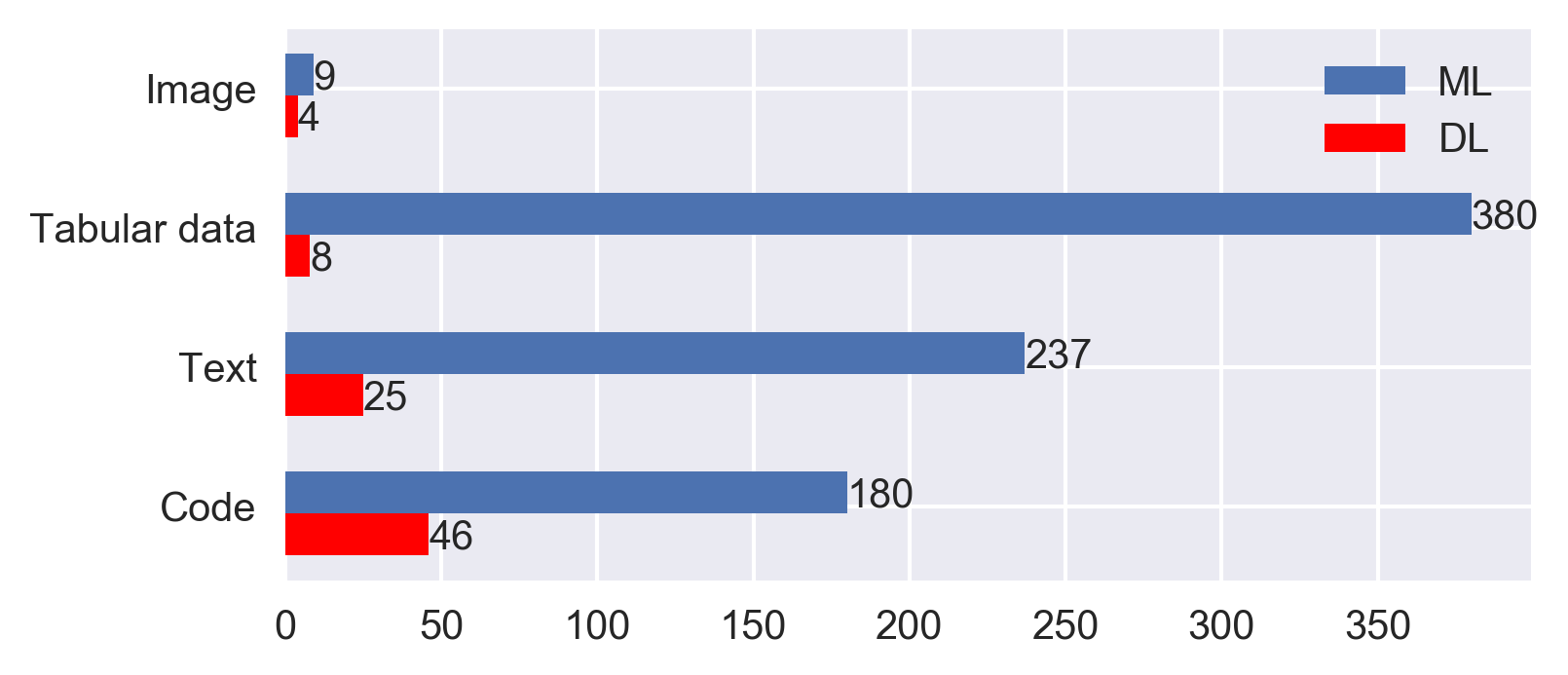}
  \caption{Distribution of ML/DL studies across four different types of data sources.}
  \label{dtype}
\end{figure}

\subsubsection{Feature Preparation}
Feature engineering is a means to exploit human ingenuity and prior knowledge to extract and organize the information from the data that is relevant to the task at hand. The performance of machine learning methods is heavily dependent on the choice of data (or feature) representations on which they are applied\cite{Bengio13}. In general, manual and automated feature engineering are the two main approaches for building ML/DL models. With regard to data, four typical types of data sources were identified in our SLR: \textit{code}, \textit{text}, \textit{tabular data}, and \textit{image}. For \textit{tabular data}, manual feature engineering is mainly applied to construct 
features %one at a time 
using domain knowledge. For the other three data sources, automated feature engineering is always applied to extract useful and meaningful features (source code, words, pixels) from datasets, yielding a framework that can be applied to any problem \cite{towardsdatascience}. 

We excluded 22 of the 891 studies that did not indicate the type of data sources and came up with a summary in Figure \ref{dtype}. We identified that 388\footnote{Some studies have more than one data source, so the numbers for each data source in Figure \ref{dtype} include the overlapping studies.} (45\%) of the remaining 869 studies used \textit{tabular data} based on manual feature engineering. However, manual feature engineering is labor-intensive and problem-dependent, and features may need to be re-designed for each new dataset \cite{Bengio13}. In order to expand the scope and ease of applicability of machine learning, it would be highly desirable to make learning algorithms less dependent on feature engineering. As a result, an increasing number of studies began to use automated feature engineering. We identified that 484 (56\%) of the 869 studies were based on automated feature engineering involving \textit{code}, \textit{text}, and \textit{Image}. 

%On the other hand, regardless of whether manual or automated feature engineering was used, one challenge of data representation concerns the difficulty in establishing a well-defined standards, or criteria for training data preparation. According to the different data sources shown in Figure \ref{dtype}, we investigated the following question: Have all these 869 studies provided detailed instructions on extracting/selecting features (or metrics) during data preparation?

Regardless of whether manual or automated feature engineering was used, a relevant question is:
\textit{\textbf{what features (or metrics) are typically extracted/selected during data preparation by SE researchers?}} 
For \textit{Tabular Data}, the software metrics/features typically used include a set of measures that provide some insights about the software system, such as the number of lines of source code, the number of classes, and the number of files. To characterize a piece of software using metrics, it is necessary to provide details about how to collect and filter metrics, thus helping choose a set of metrics with a low correlation and gain insights into the relationship among the metrics in the dataset \cite{doi:10.1002/smr.1766}. 
For \textit{Code}, SE researchers extract different syntactic information from source code as features for the ML/DL model, such as variables (declaration), methods (method signature, method body), and method calls (the API invocation sequence). For example, Wang et al.~\cite{Wang16} extracted three types of Java Abstract Syntax Tree (AST) nodes for defect prediction: nodes of method calls and class instance creations, declaration nodes, and control-flow nodes. However, since a program is defined by not only the syntax but also the semantics of the programming language used, a model for source code can be improved with the integration of the semantic information such as data types and token types, which can avoid ambiguities in the names of program and API elements (classes and method calls) and capture patterns at higher levels of abstraction \cite{8330220}.
For \textit{Text}, traditional text classification is commonly performed based on the bag-of-words (BoW) model, where n-grams or some exquisitely designed patterns are typically extracted as features \cite{AAAI159745}. Nevertheless, traditional
feature representation methods often ignore the contextual information or word order in texts and remain unsatisfactory for capturing the semantics of the words. Recent studies in SE \cite{8305930,AAAI159745,7985645,Han17} began to use pre-trained word embeddings, which can capture rich syntactic and semantic features of words in a low-dimensional vector. For example, Han et al. \cite{Han17} proposed a deep learning approach to predict the severity level of software vulnerability, using a continuous skip-gram model (one of
the popular word embedding models) to capture word-level features and a one-layer shallow CNN to capture sentence-level features in vulnerability descriptions.
For \textit{Image}, features are comprised of pixels (e.g., RGB color or grayscale pixels) that are transformed from the raw images or frames of videos. CNNs are able to maintain spatial relations between pixels by convolving the input space with a multidimensional weight matrix, commonly referred to as a filter \cite{Ott18}. Specialized image and video classification tasks often have insufficient data \cite{DBLP:journals/corr/abs-1712-04621}. Therefore, image augmentation is a solution for SE researchers, which is a technique that can be used to artificially expand the size of the training data by creating variations of the images and improve the ability of the models to generalize what they have learned to new images. %It is essential to provide detailed instructions on comparing different combinations of augmentation techniques (e.g., random rotation, shifts, shear, etc.) for a given dataset and identifying the best solution.

\subsubsection{Assumptions}
An assumption is defined as ``a thing that is accepted as true or as certain to happen, without proof'' or as ``a fact or statement taken for granted''. A software assumption refers to the software development knowledge that is taken for granted or accepted as true without evidence \cite{YANG201882}. Throughout the process of ML/DL model construction, it is important to state valid and concrete assumptions to construct models that are able to learn to predict a target output.
%which can prevent the inductive bias\footnote{The inductive bias of a learning algorithm is the set of assumptions that the learner uses to predict outputs given inputs that it has not encountered \cite{mitchell1980need}.}. %Without any assumptions, this problem cannot be solved since unseen situations might have an arbitrary output value, thus leading to inductive bias\footnote{The inductive bias of a learning algorithm is the set of assumptions that the learner uses to predict outputs given inputs that it has not encountered. \cite{mitchell1980need}}. 

Assumptions are made as the precondition in the problem definition (data preparation). For instance, to tackle the cross-version BCSD (binary code similarity detection) problem, Liu et al. \cite{10.1145/3238147.3238199} assumed that ``all binaries are compiled from source code written in high-level languages'' and ``the debug symbols in binaries are stripped''. Besides, assumptions are made in the methodology design, which are to restrict and simplify the application scenarios. For example, during the design of the ML (Case-Based Ranking) model for the problem of prioritizing software requirements, Anna et al. \cite{6249686} made the assumption in the second step (simulate preference elicitation from decision makers) that ``the simulated decision maker acts without giving inconsistent answers during the process''. This assumption restricted the
simulation to the monotonic behavior that the decision maker does not evaluate the same pair twice.

%By observing all 
Among the 891 studies, we found that assumptions for ML/DL models have several characteristics. First, it is difficult to draw the line between assumptions, requirements, risks and constraints; moreover, ML/DL studies rarely used the term ``assumption'' in their work. Second, assumptions are context dependent due to the data-driven nature of ML/DL models. Typically, an assumption could be valid for one dataset but not the other because of data changes. Third, assumptions have a dynamic nature, i.e., they can evolve over time\cite{YANG201882}. In other words, a valid assumption can turn out to be invalid at a later point of time. For example, in previous studies, defect prediction models were all built using historical data from past projects, assuming that historical data exists, the new projects share the same product/process characteristics, and their defects can be predicted based on previous data \cite{li2012sample}. However, this assumption may not always be valid because historical data may not be available (and reliable) and whether it can be used to construct effective prediction models for completely new projects remains to be verified.
%even if some previous project data is available. 
To address this problem, 
%without historical defect data,
Li et al. \cite{li2012sample} proposed a sampling-based approach to software defect prediction, only using data from current rather than past projects. Specifically, they sampled a small percentage of modules as labeled training data and used the remaining modules as unlabeled test data.

\subsubsection{Error Analysis}
Error analysis for ML/DL models is the process of examining the instances that the model misclassified, so that one can understand the underlying causes of the errors. The goals of error analysis are to (1) prioritize on which problems deserve attention and how much, (2) suggest the missing of critical information in methodology design, (3) provide a direction for handling the errors and (4) check the validity of assumptions. 

As a good example of error analysis, Liu et al. \cite{Liu18} performed a comprehensive error analysis on the misclassified pairs of changed source code and untangled change intents of AutoCILink-ML, a supporting tool for automatically identifying/recovering links between the untangled change intents in segmented commit messages and the changed source code files. They prioritized problems of the misclassified pairs. One major source of errors they found involves the misclassification of a “linked” code-intent pair (as “not linked”). 
%which leads to relatively lower precision in predicting “not linked” pairs. 
Through their error analysis, this misclassification can be attributed to the inconsistent definitions/ambiguity of certain terms used in commit messages and their related documents and the same ones used in source code. To address this kind of errors, they recommended word sense disambiguation.

%We recognized that 
Despite the benefits that an error analysis may bring, only 250 (28\%) of the 891 studies provided an error analysis or a limitation discussion for their ML/DL models. In other words, a majority of SE studies skip error analysis.

\begin{tcolorbox}[
    title = {Summary of RQ3}
    ]
We discovered that many SE studies considered performance as the top criterion for selecting ML/DL algorithms, which has two negative impacts on SE research. In addition, we found three ingredients (\textit{feature preparation}, \textit{assumptions}, \textit{error analysis}) that are essential to the understanding of the selected ML/DL techniques for a specific SE task and are quite helpful for improving the generalizability of a technique.
\end{tcolorbox}

\subsection{Analysis for DL in SE (RQ4)}
Since 2015, DL has drawn increasing attention from researchers and practitioners in the SE community due to the unique structure of DL models, which have proven effective in many SE tasks such as defect prediction, clone and similarity detection, as mentioned in Section 4.2. The number of papers is skyrocketing, especially in the past two years (2017-2018), as shown in the red bars in Figure \ref{distribution year}. With such an enormous interest, SE researchers have applied DL to 28 SE tasks across six SE activities (except \textit{Requirements Engineering}), as shown in Figure \ref{distribution SE phase}, demonstrating DL's strong capability and effectiveness in solving certain SE problems. Representation learning, the method for automatically creating a task-specific representation of the raw data, is a unique contribution of DL, as it substantially reduces the burden of feature engineering \cite{Xu:2018:PRS:3239235.3240503}. To further explore the unique characteristics and challenges of DL when integrated with SE, we identified 89 SE studies related to DL from 2009 to 2018. After excluding the 10 studies on \textit{SE for DL}, we are left with 79 studies from 25 SE tasks in our analysis below.

\subsubsection{Types of Neural Architectures Used in SE}
To enable the reader to better understand the choice of different neural architectures for different SE tasks later on, we begin by providing a brief overview of the neural architectures that are commonly used in SE. Based on the 79 studies on the applications of DL to SE problems, we observed the use of four types of deep learning architectures, namely deep neural networks (DNNs), deep belief networks (DBNs), recurrent neural networks (RNNs) and convolutional neural networks (CNNs). 

DNNs are a family of artificial neural networks with %multiple (hidden) 
one or more hidden layers between the input and output layers that aim to represent high-level abstractions in data by using model architectures with multiple non-linear transformations \cite{7961519}. A DNN is a feedforward neural network, meaning that data flows from one layer to the next without going backward, and the links between layers are one way in the forward direction. Since there are no backward links, a DNN does not have any memory: once data passes through the network, it is forgotten, and it cannot be exploited (as historical context) to predict data items that are encountered at a later point in time. A DBN is a generative graphical model that uses a multi-level neural network to learn a hierarchical representation from training data that could reconstruct the semantic and content of input data with a high probability \cite{Wang16}. A CNN is especially well suited for image recognition and video analysis tasks  because a CNN, which is inspired by the biological findings that the mammal’s visual cortex has small regions of cells that are sensitive to specific features of the visual receptive field \cite{Chen:2018:UDI:3180155.3180240}, can exploit the strong spatially local correlation present in images. A RNN, unlike a DNN, allows backward links and therefore can remember things. It is therefore well suited for processing sequential data such as text and audio because its feedback mechanism simulates a "memory" so that a RNN’s output is determined by both its current and prior inputs \cite{7985645}. While a RNN has memory, its memory may not be that good. More specifically, it may only remember things in the recent past and not those that it saw a while ago due to a problem known as vanishing gradient. To address this problem with the standard RNN model, two variants are widely adopted in SE with mechanisms for capturing long-term dependencies: Long Short Term Memory (LSTM) networks and Gated Recurrent Units (GRUs).

\begin{figure}[t]
  \centering
  \includegraphics[scale = 0.8]{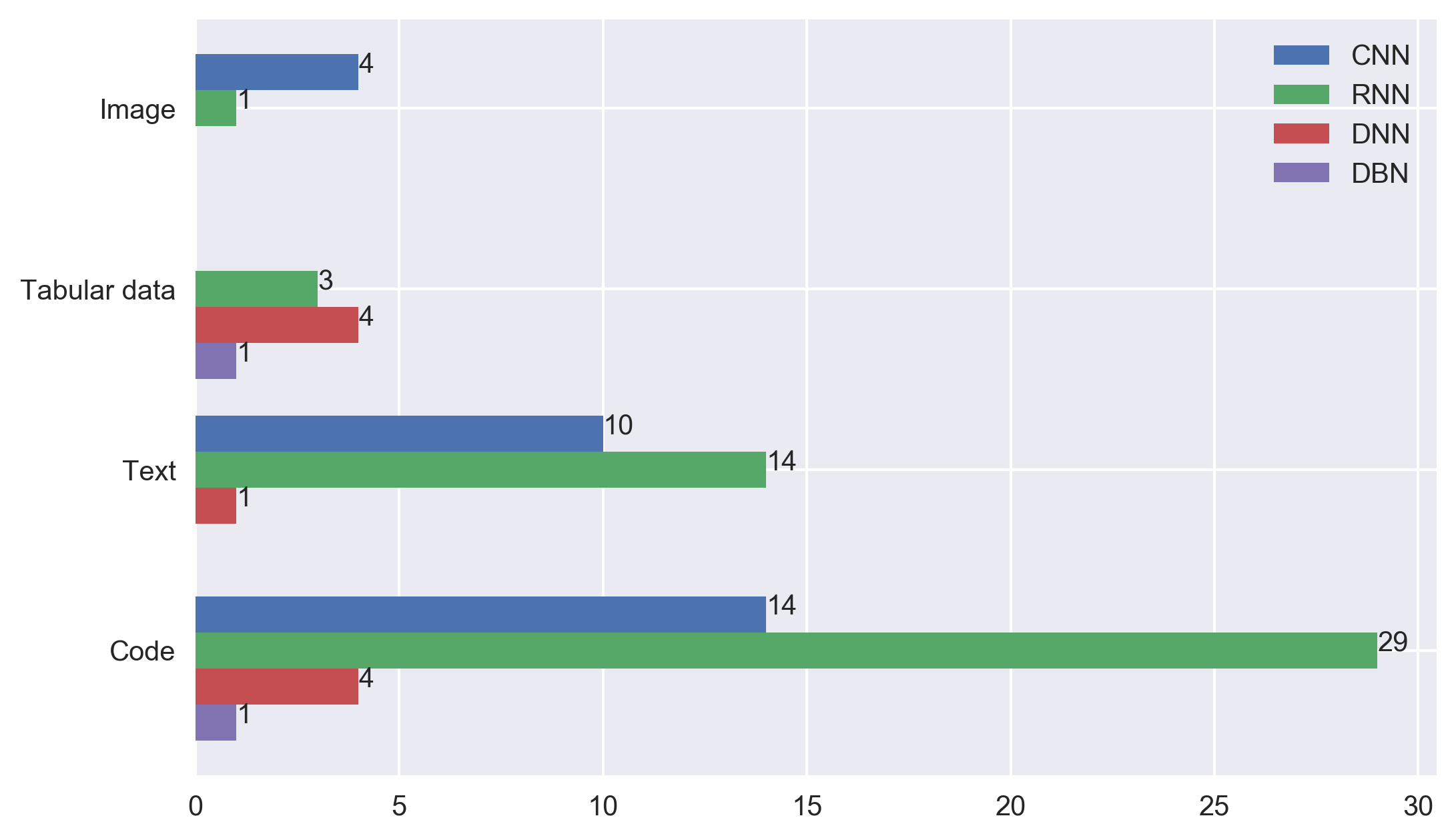}
  \caption{Distribution of DL studies with four different DL architectures across four different types of data sources.}
  \label{dltype}
\end{figure}

\textit{\textbf{How are various SE data sources and tasks related to different types of DL models?}} Figure \ref{dltype} shows how the different types of neural architectures used in DL-based studies in SE are distributed over four different data sources.
%that  different types of neural architectures DL studies spread over all four different data sources. A further analysis %(see Figure \ref{dltype}) 
A closer inspection revealed that the data sources do not have a direct relation with various types of DL models because at least two different types of DL models or  hybrid models (e.g., CNN+RNN) are applied to each data source. 

Given the above discussion, it is perhaps not surprising that there is no one-to-one mapping from data sources to neural architectures. The reason, again, is that the choice of a neural architecture for an SE task depends not only on the data source, but also on how we want to {\em encode} the input. If we want to encode the spatial locality in the input, then a CNN would be a good choice. For instance, the pixels in an image that are spatially close to each other are more likely to be related to each other (e.g., they may belong to the same object) than those that are far apart from each other. As another example, the words that are close to each other in a sentence are more likely to be related to each other (e.g., they may belong to the same linguistic construct, such as a noun phrase or a verb phrase) than those that are far apart from each other. A CNN allows us to capture spatial locality by applying {\em filters} of different sizes to extract high-level features from elements that are spatially close to each other via an operation known as {\em convolution}. In contrast, if we want to encode sequences (e.g., to capture word order in a text document), then a RNN such as a LSTM would be a good choice: as mentioned above, LSTMs allow the elements of a sequence to be processed sequentially so that the elements that appear later in the sequence can exploit information from elements that appear earlier in the same sequence. Finally, if we do not care about exploiting spatial locality or temporal correlations, then we can always apply a DNN to a given task: in a DNN, the features extracted from the input are treated as independent of each other. Hence, a DNN is like the majority of the traditional machine learning algorithms in the sense that features are assumed to be independent and that they are typically generic enough to be applicable to any dataset.

Consequently, some model types are preferred to others for a given data type.
%the choice of a neural architecture for a given task is less random than the choice of a traditional machine learner. 
In fact, we can see from Figure~\ref{dltype} that the distributions of models over different data sources are not entirely random. For instance, CNNs are more often used for processing images because of the need to capture spatially local correlations in images, whereas CNNs and RNNs are both frequently used for processing because of the need to encode spatial locality and temporal constraints. We also see that RNNs are more popularly used than CNNs for code processing. We will return to this observation later in this section.  

For those readers who are not familiar with DL, we note that these learners are “deep” because there can be multiple layers in the network between the input layer and the output layer, making the processing pipeline “deep”. As mentioned before, a CNN/RNN {\em encodes} the input as some form of {\em representation}. We can {\em stack} CNNs and RNNs. In other words, we can apply multiple filters and multiple LSTMs, for instance, where later filters/LSTMs will produce increasingly abstract representations of the input. After multiple layers of encoding, the resulting representation is typically passed to one or more so-called {\em dense} layers, which are typical feedforward networks, before reaching the output layer. Each node is associated with a non-linear activation function. Hence, a network with numerous layers of non-linear units will be able to express highly non-linear, arbitrarily complex functions.

Note that the neural architectures we have described so far are by no means new: these are technologies developed in the 1980s. However, DL does not become popular until recently because they are data-intensive. More specifically, because of the large number of layers in a neural network, the number of parameters is large, and without a “sufficiently large” training set, the network can easily overfit the training data. As noted in the introduction, the success and popularity of DL in the computer vision community was triggered by the development of ImageNet, a large-scale data resource for image recognition that has enabled the use of DL, not by the development of new DL technologies. With sufficient training data, deep learners can easily outperform traditional machine learners because the representations they learn can typically encode important features of the input that traditional learners fail to encode. This is precisely what makes DL appealing to researchers in different communities.

\subsubsection{Recent Major Developments in DL}
As mentioned above, the neural technologies we described above are not new. In this subsection, we provided a high-level overview of some of the exciting recent developments in DL that we believe would be of interest to SE researchers, and which may inspire new ideas as we present the existing applications of DL to SE in the next subsection.

{\bf The Encoder-Decoder Framework.} One of the most important recent developments in DL is arguably the encoder-decoder framework, which is fairly well-known in the SE community. This framework enables so-called sequence-to-sequence ({\sc Seq2Seq}) learning \cite{sutskever:nips14}. As the name suggests, this neural architecture takes a sequence as input and produces a sequence as output. Hence, it is a natural framework for tasks such as machine translation, which takes as input a word sequence in the source language and outputs a word sequence in the target language, as well as summarization, where the input is a sequence of words or code elements and the output is a textual summary (i.e., a word sequence).

How is the encoder-decoder framework related to the neural architectures we described in the previous subsection? Recall that these architectures are {\em encoders}: they encode a given input as a vector representation that is then used for some prediction task. In our case, we want to predict a sequence as the output. In other words, we want the neural model to help us map the vector representation produced by the encoder to an output sequence. This is done using a {\em decoder}: in each timestep, the decoder decodes/outputs one element in the output sequence based on the vector representation and all the elements it has decoded so far up till the current timestep. Since the elements are generated sequentially and the later elements are dependent on their earlier counterparts, a RNN is a natural choice for decoding. This explains why RNNs are a popular choice for code-related tasks (e.g., code summarization) in Figure~\ref{dltype}. Note that the choice of the encoder and that of the decoder do not have to be dependent on each other in the encoder-decoder framework. For instance, a CNN can be used for encoding and a RNN can be used for decoding. 

The encoder-decoder framework has enabled significant advances in summarization and generation tasks in the SE community. To understand its significance, recall that traditional approaches to summarization/generation proceed in multiple steps, where one needs to first extract the relevant information from the input, and then the extracted elements are fed into some hand-crafted templates for generating the output. In contrast, {\sc Seq2Seq} learning provides a so-called end-to-end framework where a given input is being mapped to the desired output directly in a single model. 

{\bf The Attention Mechanism.} Attention is a mechanism designed to improve the encoder-decoder framework \cite{bahdanau:iclr15}. As mentioned above, the decoder takes the output of the encoder and what has been decoded so far to generate the next element in the output sequence. Typically, the hidden state of the last timestep of the encoder is used as its output. However, there have been concerns over whether merely using the last hidden state of the encoder can yield an adequate representation of the input. 
Attention is designed to address this problem. When generating an element in the output sequence in a given timestep, some elements in the input sequence may have a greater influence than the other. Attention helps identify the relevant/influential elements in the input sequence with respect to a particular output element and encode them in a vector known as the {\em context} vector, which will be used as an additional input to the decoder. More specifically, to generate an output element using the decoder, attention is first used to compute a weight for each element in the input sequence indicating its importance with respect to the output element to be generated, and the resulting weight distribution is then used to compute the aforementioned context vector. Intuitively, the context vector amplifies the relevant information from the input and de-emphasizes the not-so-relevant information from the input. Attention has been shown to be effective in improving the encoder-decoder framework and has been extensively used by SE researchers. %Attention has recently been extensively used by SE researchers in combination with the encoder-decoder framework.

{\bf Word Embeddings.} 
%When the input to a deep neural network is an image, the image is typically encoded as a 2-dimensional array of pixels. 
When the input is a text document, the document is typically encoded as a sequence of words. The words, however, are typically represented as a word vector. In traditional machine learning, a word is encoded as a “one-hot” vector, where the length of the vector is the number of words in the vocabulary, and the value of an element is 0 except for the element corresponding to the word itself, which has a value that is either 1 or its tf-idf. One-hot vectors, however, are a poor representation of word semantics, because two semantically similar words are not more similar to each other than two semantically unrelated words in the one-hot representation. To address this problem, researchers have trained word {\em embeddings}. Unlike one-hot vectors, which are sparse representations of words, word embeddings are dense, distributed vector representations of words that are trained such that the embeddings of two semantically similar words (e.g., "delighted" and "glad") are more similar to each other than the embeddings of two semantically unrelated words (e.g., "happy" and "tired"). In other words, word embeddings capture word semantics much better than traditional one-hot vectors. Perhaps not surprisingly, because word embeddings are better at capturing word semantics, they also enable a neural model to learn a better representation of the input (i.e., a representation that can encode the underlying meaning of the input). The good thing about word vectors is that they can be trained in an unsupervised manner on large text corpora. Word embeddings are getting better over the years. For instance, in first-generation word embeddings produced by Mikolov et al. \cite{mikolov:nips13} using a continuous skip-gram model, the embeddings are context-independent, meaning that a given word, regardless of the context in which it appears, will have exactly one embedding. Newer embeddings are {\em contextualized} \cite{peters:naacl18}, meaning that a given word can have multiple embeddings depending on the context in which it appears. While word embeddings have been used to train neural models in the SE community, contextualized embeddings have not even though they can potentially take DL-based SE systems to the next level of performance.  

{\bf New Encoders.} While CNNs and LSTMs/GRUs are natural choices for encoding images and sequences respectively, there are data instances that are structurally arguably more complex than images and sequences, such as trees and graphs. Trees and graphs are data structures that are commonly found in SE (e.g., control flow graphs, API call graphs, abstract syntax trees). To exploit the inherent dependencies between the elements in these data structures, researchers have specifically designed new neural models such as tree LSTMs \cite{tai:acl15} and graph convolutional neural networks \cite{bruna:iclr14} for embedding different kinds of inputs.  

{\bf Pretrained Models.} When we apply DL, we would go through the typical process of creating annotated training data to train a neural model. However, it is no secret that manual annotation of training data is a time-consuming and labor-intensive process, and the fact that neural models are data-intensive only aggravates the situation. {\em Pretrained models} are a recent promising idea for addressing this problem. Assume, for instance, that we are interested in a task for which only a small amount of annotated data is available for training, but we may have plenty of training data for a related task.
%that would allow us to train a neural model. 
In this case, we can first train a neural model on this related task. This model will become a {\em pretrained} model for the task in which we are interested: the weights stored in this pretrained model may contain knowledge that would benefit our task. Using this pretrained model as a starting point, we can {\em fine-tune} this model by training it on the small amount of labeled data available for our task so that it can be adapted to the task at hand. In other words, rather than training a model from scratch, we train a model that has been pretrained on a related task, which would allow us to transfer knowledge from the related task to our task (via so-called {\em transfer learning}). Because of this ability to exploit knowledge from a related task, we may be able to reduce the amount of annotated training data needed for our task to reach a given level of performance. Key to the success of transfer learning would be to identify tasks that are related to each other. Several well-known pretrained models, such as Transformer \cite{Vaswani:nips17} and BERT \cite{devlin:naacl19}, have been developed in the last few years in the text processing community. These models have been pretrained on a variety of text processing tasks, so they are believed to possess general linguistic knowledge that would benefit a variety of language processing tasks. To our knowledge, pre-training is an idea that is under-investigated in the SE community and certainly deserves attention.

\begin{table*}[htbp]
\scriptsize
  \renewcommand{\arraystretch}{2}
  \caption{Summary for DL Studies.}
  \label{tab:summary_for_DL}
  \begin{tabular}{p{1.5 cm} p{3.8 cm} p{4.5cm} p{5cm}}
    \hline
    \textbf{Impact Level} & \textbf{SE Task} & \textbf{DL Approaches} & \textbf{DL-Enhanced Approaches}\\
    \hline
        \multirow{2}{2 cm}{New} & Code Summarization & RNN / CNN Encoder-Decoder Model \cite{iyer-etal-2016-summarizing,Hu:2018:DCC:3196321.3196334,Jiang:2017:AGC:3155562.3155583,pmlr-v48-allamanis16} & 1. Deep Reinforcement Learning \cite{10.1145/3238147.3238206} \newline 2. Code-GRU Model \cite{liang2018automatic}\\ 
        
        & Source Code Identification & CNN Model \cite{Ott18,Alahmadi:2018:APL:3273934.3273935} & N/A\\
        
        \hline
        
        \multirow{18}{2 cm}{Transition} & API Mining & RNN Encoder-Decoder Model \cite{10.1145/2950290.2950334} & N/A\\
        
        & Traceability Link Generation and Recovering & Bidirectional GRU Model\cite{7985645}  & N/A\\
        
        & Specification Mining & RNN Model \cite{10.1145/3213846.3213876} & N/A\\
        
        & Code Clone and Similarity Detection & 1.DNN Model \cite{10.1145/3236024.3236026,8094426,10.1145/3236024.3236068} \newline 2.RNN Model \cite{White:2016:DLC:2970276.2970326,8595238} \newline 3.CNN Model \cite{10.1145/3238147.3238199,8530727} & N/A\\
        
        & Data and Service Modeling & DNN Model \cite{MARRON2017195} & N/A\\
        
        & Model Optimization & CNN Model \cite{li2018deeprebirth} & N/A\\
        
        & Program Synthesis & RNN Model \cite{10.1145/3236024.3236066} & N/A\\
        
        & Code Smell/Anti-pattern Detection & CNN Model \cite{8330265,10.1145/3238147.3238166} & N/A\\
        
        & Test Case Generation & RNN Model \cite{8115618,7985701,10.1145/3213846.3213848} & N/A\\
        
        & Test Automation & Deep Reinforcement Learning \cite{8529836} & N/A\\
        
        & Repository Mining & 1.CNN Model \cite{Menzies18,7582810,Xu:2018:PRS:3239235.3240503,Fu:2017:EOH:3106237.3106256,Xu:2016:PSL:2970276.2970357} \newline 2. LSTM Model \cite{labutov-etal-2018-multi} & N/A\\
        
        & Defect Fixing & RNN Model \cite{8453063,8330219} & N/A\\
        
        & Software Application/Artifact Categorization & 1.CNN+LSTM Model\cite{8530052} \newline 2.CNN Model \cite{doi:10.1142/S1469026818500189} & N/A\\
        
        & Bug Report and Review Summarization & Stepped Auto-encoder Network \cite{Li:2018:UDB:3196321.3196326} & N/A\\
        
        & Software Quality Prediction & 1.CNN Model \cite{Han17,MI201860} \newline 2.CNN+LSTM Model \cite{YAN201867} & N/A\\
        
        & Schedule/Effort/Cost Estimation & DNN Model \cite{8009938} & N/A\\
        
        & Performance Prediction & RNN Model \cite{8009938Hashemi} & N/A\\
        
        & Energy Estimation & LSTM Model \cite{8094428} & N/A\\
        
        \hline
        
        \multirow{5}{2 cm}{Enhanced} & Defect Prediction & 1.DBN Model \cite{Wang16,7272910} \newline 2.RNN/LSTM Model \cite{8330212,8424963,10.1145/3236024.3236060} \newline 3.CNN Model \cite{Li17} & Graph-based CNN Model \cite{8371922}\\
        
        & Defect Detection and Localization & 1.DNN Model \cite{7961519,10.1145/3238147.3238163}\newline 2.CNN Model \cite{Kan18,10.1145/3236024.3236082,8094414} & 1.CNN+Random Forest Model \cite{10.1145/3210459.3210469} \newline 2.Enhanced CNN Model \cite{8305956}\\
        
        & Code Optimization & 1.RNN Model \cite{White:2015:TDL:2820518.2820559,Hellendoorn:2017:DNN:3106237.3106290,10.1145/3236024.3236051,duong2017multilingual} \newline 2.Deep Feedforward Neural Network \cite{xu2017local} & 1.Multi-prototype DNN Language Model \cite{8330220} \newline 2. Code-Description Embedding Neural Network \cite{8453172} \newline 3.Hybrid Code Networks \cite{williams2017hybrid}\\
        
        & Code Generation and Completion & 1.RNN Encoder-Decoder Model \cite{8330222,yin-neubig-2017-syntactic} \newline 2.LSTM Model \cite{parvez-etal-2018-building} \newline
        3.RNN+CNN Model \cite{Chen:2018:UDI:3180155.3180240,gupta-etal-2018-uncovering} & 1.Two-step Attentional Encoder-Decoder Model \cite{iyer-etal-2018-mapping} \newline 2.Latent Prediction Network \cite{ling-etal-2016-latent}\\
        
        & Sentiment Analysis & RNN Model \cite{AAAI148148,10.1145/3180155.3180195} & 1.Hierarchical Multimodal GRU Model \cite{gu-etal-2018-multimodal} \newline 2.Recurrent Convolutional Neural Networks \cite{AAAI159745}\\
        
  \hline
\end{tabular}
\end{table*}

\subsubsection{Impact Analysis for DL Studies}

\textit{\textbf{What impact does DL bring to SE tasks?}} According to the impact level of DL for SE tasks, as shown in Table \ref{tab:summary_for_DL}, we classified all 79 DL studies into three categories: New (No effective ML methods before), Transition (from ML-based to DL-based approaches), Enhanced (Enhanced DL approaches emerged after ML and basic DL approaches). 

For the ``new'' category, as mentioned in Section 4.2.1, the emerging DL techniques addressed the problems for two SE tasks (\textit{Source Code Identification} and \textit{Code Summarization}) that ML techniques were not capable of tackling. \textit{Source Code Identification} refers to a task of automatically extracting correct code appearing in video tutorials \cite{Alahmadi:2018:APL:3273934.3273935}. Existing approaches by using Optical Character Recognition (OCR) techniques often mix the non-code text with the source code, resulting in incorrect and unusable code being extracted \cite{Alahmadi:2018:APL:3273934.3273935}. Therefore, it is necessary to first accurately identify the section of the screen where the code is located and then apply OCR only to that section. With the powerful and accurate object recognition abilities through the use of filters, CNNs are currently the best choice to  classify the presence or absence of code \cite{Ott18} and predict the exact location of source code within each frame \cite{Alahmadi:2018:APL:3273934.3273935}, which accelerates code identification and code example resolution in video tutorials. 
\textit{Code Summarization} refers to a task of comprehending code and automatically generating descriptions directly from the source code \cite{10.1145/3238147.3238206}, including commit messages and code comments. Most of the existing code summarization methods learn the semantic representation of source codes based on statistical language models. However, a statistical language model (e.g., the n-gram model) has a major limitation: it predicts a word based on a fixed number of predecessor words \cite{10.1145/3238147.3238206}. Following the trend of employing the DL-based attentional encoder-decoder framework, recent studies have built a language model for natural language text and aligned the words in text with individual code tokens directly using an attention component. These DL studies can predict a word using preceding words that are farther away from it. In addition, two enhanced DL approaches were introduced that improve performance by around 20\% and 10\% respectively in terms of ROUGE\footnote{The ROUGE score counts the number of overlapping units between a generated sentence and a target sentence \cite{liang2018automatic}.}. First, Wan et al. \cite{10.1145/3238147.3238206} integrated reinforcement learning into the attentional encoder-decoder framework to solve the biased assumption that decoders are trained to predict the next word by maximizing the likelihood of next ground-truth word given the previous ground-truth word. Using deep reinforcement learning, one can generate text from scratch without relying on ground truth in the testing phase. Second, due to the fact that the attentional encoder-decoder framework does not exploit the code block representation vectors, Liang et al. \cite{liang2018automatic} proposed a new RNN model called Code-RNN, which gets a vector representation of each code block and this vector contains rich semantics of the code block.

For the ``Transition'' category, because of DL's ability to represent data features with multiple levels of abstraction and its proven effectiveness when applying to source code, researchers in 18 SE tasks have begun to transit from ML to DL methods. Four typical examples are presented below, including \textit{API Mining}, \textit{Specification Mining}, \textit{Traceability Link Generation and Recovering} and \textit{Code Clone and Similarity Detection}. First, Gu et al.  \cite{10.1145/2950290.2950334} successfully adopted RNN Encoder-Decoder (DeepAPI) to \textit{API Mining} that generates API usage sequences for a given natural language query. DeepAPI learns the sequence of words in a query and the sequence of associated APIs, which can produce more accurate API usage sequences compared to state-of-the-art techniques and the improvement was over 170\% in terms of BLEU score. Second, Le et al. \cite{10.1145/3213846.3213876} used DL techniques for \textit{Specification Mining}, which help developers reduce the cost of manually drafting formal specifications. They employed a Recurrent Neural Network Based Language Model (RNNLM) on the collected traces to predict the next likely method to be executed given a sequence of previously called methods. The RNNLM outperformed all existing specification mining methods by around 30\% in terms of average F-measure. Third, Guo et al. \cite{7985645} introduced DL techniques for \textit{Traceability Link Generation and Recovering}, the task of automatically creating, maintaining and recovering trace links among various kinds of software artifacts. They utilized word embeddings and RNN models to predict the likelihood of a trace link between two software artifacts, significantly improving the trace link accuracy in comparison to standard baseline techniques by 30\% to 40\% in terms of Mean Average Precision (MAP). Fourth, as mentioned in Section 4.2.1, several studies have begun to leverage DL techniques for \textit{Code Clone and Similarity Detection}, which involve measuring code similarity and detecting four types of code clones that are common in software systems. For instance, Li et al. \cite{8094426} implemented the first solely token-based clone detection approach using a DNN, which effectively captured the similar token usage patterns of clones in the training data and detected nearly 20\% more Strong Type 3 clones than ML approaches.

For the ``Enhanced'' category, while the studies from \textit{Defect Prediction}, \textit{Defect Detection and Localization}, \textit{Code Optimization}, \textit{Code Generation and Completion} and \textit{Sentiment Analysis} have previously utilized ML-based approaches and simple DL-based approaches, they currently involved some continuously enhanced DL approaches. Two typical examples are illustrated below. For \textit{Defect Prediction}, most existing approaches started by exploiting the tree representations of programs, typically the Abstract Syntax Trees (ASTs). They simply represented the abstract syntactic structure of source code but did not show the execution process of programs, so software metrics and AST features may not reveal many types of defects in programs. Phan et al. \cite{8371922} formulated a directed graph-based convolutional neural network (DGCNN) over control flow graphs (CFGs) that indicate the step-by-step execution process of programs to automatically learn defect features. DGCNNs can treat large-scale graphs and process the complex information of vertices like CFGs, significantly outperforming baselines by 1.2\% to 12.39\% in terms of the accuracy. For \textit{Defect Detection and Localization}, two enhanced DL approaches were proposed to improve the MAP by around 5\%. First, there is a dilemma for current bug localization techniques: ML approaches ignored the semantic information between the texts in bug reports and code tokens in source files, while %existing 
DL approaches ignored the structural information of both bug reports and source files \cite{10.1145/3210459.3210469}. Motivated by this observation, Xiao et al. \cite{10.1145/3210459.3210469} proposed CNN\_Forest, a CNN and Random Forest-based approach where an ensemble of random forests is applied to detect the structural information from the source code and the alternate cascade forest works as the layer-structure in the CNN to learn the correlated relationships between bug reports and source files. Second, current studies using DL achieved poor performance and most improvements still came from Information Retrieval (IR) techniques (which focused more on textual similarity than semantics). In other words, the final results may still be heavily influenced by the performance of IR \cite{8305956}, meaning that the deep neural network in their model was more like a subsidiary. Xiao et al. proposed an enhanced model, DeepLocator, for bug localization, which consists of a revised TF-IDuF (term frequency-user focused inverse document frequency) method, word2vec and an enhanced CNN by adding bug-fixing recency and frequency in the fully connected layer as two penalty terms to the cost function. DeepLocator correlated the bug reports to the corresponding buggy files rather than relying on the textual similarity used in IR-based approaches. 

\begin{tcolorbox}[
    title = {Summary of impacts for RQ4}
    ]
Based on the 79 DL studies classified by 25 SE tasks, we identified two SE tasks where emerging DL techniques initially addressed the problems which ML techniques were not capable of tackling, 18 SE tasks for which ML methods are gradually superseded by their DL counterparts, 
%methods started being explored following the existing ML methods, 
and five SE tasks where DL approaches sought continuous enhancements for further improvement.
\end{tcolorbox}

%\textit{\textbf{How are various SE data sources related to different types of DL models?}} Figure \ref{dtype} shows that DL studies spread over all four different data sources. A further analysis (see Figure \ref{dltype}) reveals that data sources do not have a direct relation on various DL models because at least two different types of DL models or a hybrid (i.e., CNN+RNN) model are applied to each data source. Nevertheless, two observations deserve mention. First, DBN is rarely used in SE due to the fact that DBN has mostly fallen out of favor, even when compared to other unsupervised or generative learning algorithms \cite{goodfellow2016deep}. Second, CNN is still the top choice for solving image classification problems because CNN, which is inspired by the biological findings that the mammal’s visual cortex has small regions of cells that are sensitive to specific features of the visual receptive field \cite{Chen:2018:UDI:3180155.3180240}, can exploit the strong spatially local correlation present in images.

\begin{table}[t]
\footnotesize
  \renewcommand{\arraystretch}{1.5}
  \caption{Data Size of ML- and DL-related studies in SE.}
  \label{tab:Data_Size}
  \begin{center}
  \begin{tabular}{p{0.5cm} p{3cm} p{1cm} p{2cm} p{2cm}}
    \hline
     & \multirow{2}{3cm}{\textbf{Median Data Size}} & \multicolumn{3}{c}{Paper Count}\\\cline{3-5}
     & & \textbf{Total} & \textbf{size$\geq 1000 $} & \textbf{size$\geq 10000$}\\
    \hline
        ML & 7842 & 781 & 633 (81\%) & 366 (47\%)\\
        \hline
        DL & 51831 & 75 & 71 (95\%) & 57 (76\%)\\
  \hline
\end{tabular}
\end{center}
\end{table}

\subsubsection{Challenges for DL in SE}
Compared to traditional ML models, DL models typically require a larger amount of data to train.
%and test. 
Table \ref{tab:Data_Size} shows a comparison of the data size (number of data entries) used for training and testing by ML and DL studies in SE. Excluding the studies that did not explicit report data size (four of the 79 DL studies and 31 of the 812 ML studies respectively), we found that 57 (76\%) of the 75 DL-related studies have been trained and tested on data with over 10000 entries (see Table \ref{tab:Data_Size}). \textbf{The large data size, together with the inherent complexity of DL models, impose two big challenges in training and testing DL models for SE tasks.} Specifically, a significant amount of manual effort and expertise was needed for data labeling and annotation, and significant time and computational resources were required to train DL models. Furthermore, we observed that 45 (57\%) of the 79 DL studies considered performance as the top criterion for selecting DL algorithms but overlooked the applicability and generalizability of the proposed DL-based approaches. This could limit the %potential 
contributions of \textit{DL for SE}. One example that is worth noting is that \textbf{due to the sheer number of parameters and the complex structure, there is still no effective way to interpret the internal representation of DL models}. Given the above observations, we will discuss five challenges that need to be addressed to better leverage DL to improve the productivity of SE tasks:
%three bottlenecks which may hinder the synergy between DL and SE and project future directions for research.

%\textit{\textbf{Future directions for DL research.}} We suggest three directions for DL research in SE:
%\begin{itemize}
    %\item \textit
    {\bf Reducing manual efforts in data labeling and annotation.} Owing to the diversity of SE tasks, publicly available annotated datasets or benchmarks are scarce for SE researchers and practitioners. Manually annotating over a thousand or even ten thousand samples is always a mandatory process for DL applications when given a completely new problem \cite{Alahmadi:2018:APL:3273934.3273935}. Ensuring high quality data labeling and annotation requires not only a significant amount of manual effort but also valuable domain expertise in various SE tasks. Researchers have started to address this issue. Unsupervised DL, which does not require manually labeled training data, was employed in a limited number of studies when labeled data is lacking. For instance, Li et al. \cite{Li:2018:UDB:3196321.3196326} proposed a novel unsupervised algorithm for bug report summarization, which assigned weights of words and sentences without human mined features. However, compared to supervised techniques, its performance is unstable. Recently, a crowdsourcing approach was attempted by Ott et al. \cite{Ott18} to acquire labeled data, but it still needed extra effort to evaluate the quality of the resulting data. Semi-supervised techniques, which use clustering techniques with a small number of labeled instances, can be another potential solution. It has been adopted in some ML applications in SE \cite{Kaleeswaran:2016:SVF:2950290.2950363, Jing:2016:MDI:2884781.2884827}, but it is still rarely seen in DL.
    
    {\em Active learning} should be considered when it comes to reducing manual annotation effort. Recall that active learning aims to reduce annotation effort by selecting only the informative examples in the training set for manual annotation in an iterative fashion \cite{cohn:ml94}, so as to minimize the number of manually annotated examples needed to reach a certain level of performance. We believe that combining DL with active learning is an unexplored, but promising avenue of research for SE researchers.

Finally, pretrained models (i.e., the use of a neural model pretrained on one or more related tasks) may be a promising way to reduce manual annotation effort in SE. As mentioned before, key to the success of using pretrained models for transfer learning would be to identify tasks that are related to each other. An idea closely related to transfer learning is {\em multi-task learning}. Assume that we are interested in a task for which only a small amount of annotated data is available, and at the same time there is a related task for which we also have a small amount of annotated data. In this scenario, we can employ a multi-task learning framework where we train both tasks in the same network simultaneously. In other words, the network will have two outputs (one for each task), but by training both tasks in the same network, they can share the same representation, meaning that information learned from one task can be exploited when learning the other task. We believe that transfer learning and multi-task learning are both promising avenues for solving the data paucity problem that are under-investigated in the SE community.

{\bf Deep learning with feature engineering.} While one of the often cited advantages of DL is that it obviates the need to perform manual feature engineering because of the ability of DL models to automatically learn task-specific representations, an under-investigated question is whether the performance of DL models can further be improved with the incorporation of hand-crafted features. One may wonder why a DL model cannot learn/infer such features from the input. One possibility is that a task may require background knowledge that is not explicitly available in the input. In that case, merely learning representations of the input will not provide sufficient knowledge for addressing the task. Another benefit of hand-crafted features is that they might reduce the amount of labeled training data needed to reach a certain level of performance. The reason is that hand-crafted features may encode a human’s knowledge of what would be useful for the task at hand that may otherwise have to be automatically acquired from labeled training data.
%which, when absent, will need to be automatically acquire from a potentially larger amount of labeled training data. 

    %\item \textit
    {\bf Evaluating and reducing computation cost.} Note that among the 79 studies, 69 (87\%) did not reported their runtime,
    %in evaluating their computation cost, 
    which may prevent the follow-on studies from comparing the computational costs with theirs and make it difficult to transfer the experimental results into real-world practices. The computational challenge in training DL models has been recognized and considered as a potential threat of generality in SE \cite{Li:2018:UDB:3196321.3196326, Fu:2017:EOH:3106237.3106256, Hellendoorn:2017:DNN:3106237.3106290}. Two ways have been suggested to reduce the computation time: (1) making software analytics faster and (2) exploiting hardware features. For instance, local learning can divide training data into small pieces and then one model can be learned from each piece \cite{Menzies18}. Alternatively, cloud computing is an ideal choice for designing, developing and training deep learning applications efficiently. It allows large datasets to be easily ingested and managed, and algorithms can be trained by leveraging distributed networks. %More
    Scalable optimization methods for SE data are expected to scale the training of DL models from a single machine to as many machines as possible, such as synchronous and asynchronous stochastic gradient descent, and taking advantage of multiple processors, distributing workloads seamlessly and efficiently across different processor types and quantities \cite{DBLP:journals/corr/JinYIK16}.
    %\item \textit
    
    {\bf Gaining SE insights from DL.} We found that very few DL studies interpreted the features learned from deep neural networks and explained how the learned DL models could be applied to real-world applications. Hence, in addition to the breakthroughs on model performance, we expect future research to improve the interpretability of neural network representations \cite{Zhang2018}. The difficulty in interpretability has always been a limiting factor for use cases requiring explanations of the task-specific features  to transit from manual to automated process in SE tasks. Interpretability is a hot topic in AI. We anticipate that future research could bring additional insights into how improved DL model performance can be turned into more effective and efficient SE practices and what changes in SE practices would be useful to optimize the proposed DL-based approach. To maintain a healthy eco system for the synergy between DL and SE, we encourage researchers to evaluate their DL-based approaches on additional metrics beyond performance in order to enhance the applicability and generalizability of the approaches.
%\end{itemize}

{\bf Learning SE-specific embeddings.} As mentioned before, the increasingly successful application of DL to text has been enabled in part by the development of better and better word embeddings. Most importantly, these word embeddings are pretrained on large, unannotated corpora. A promising direction for SE researchers would be to train embeddings for the types of data that are typically encountered in SE, such as code. Code, like text, is sequential, so we expect that embeddings of code elements (e.g., API) can also be trained in a similar, unsupervised manner as words. Using such embeddings has the potential to take the performance of various SE tasks to the next level, such as API recommendation.

\begin{tcolorbox}[
    title = {Summary of challenges for RQ4}
    ]
We discovered five unique challenges to be addressed to better leverage DL to improve the productivity of SE tasks: (1) a significant amount of manual effort and expertise was needed for data labeling and annotation, (2) the integration of feature engineering with DL models, (3) significant time and computational resources were required, (4) the lack of an effective method for interpreting the internal representations of DL models, and (5) the development of SE-specific embeddings.
\end{tcolorbox}

\begin{figure*}[t]
  \centering
  \includegraphics[width=\linewidth]{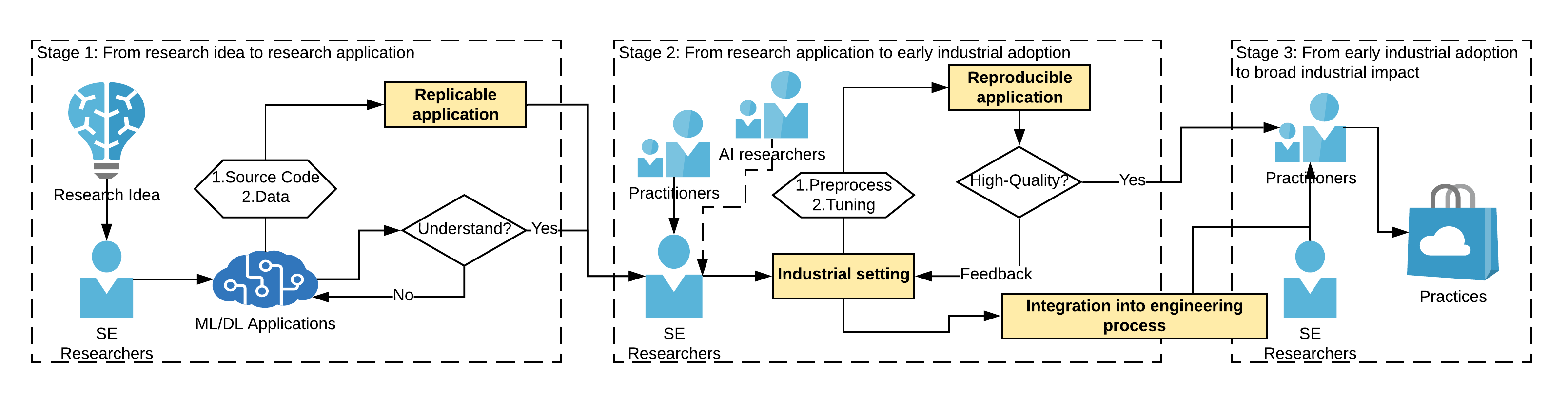}
  \caption{A road-map to industry practice.}
  \label{roadmap}
\end{figure*}

\section{Discussion}
\subsection{Transferring ML/DL Research Results to Real-World Applications}
The ultimate goal for SE researchers is to transfer ML/DL-related research results into real-world SE practices. Our SLR results show that although ML/DL or their combination has been integrated with 60 SE tasks, only 61 (7\%) of the 906 studies mentioned their collaborations with industrial partners or tested their approaches with industrial data and settings, as mentioned in Section 4.3. This suggests the transfer of ML/DL applications from academia to industry still has a long way to go. Significant efforts are needed from both SE researchers and industrial practitioners. Based on a successful case described in \cite{Dang17} as well as our findings in Section 4, we propose a three-stage road-map (Figure \ref{roadmap}) that SE researchers could employ to turn ML/DL-related research ideas into real-world applications.

The first stage involves demonstrating and validating contributions of the created ML/DL applications in lab settings. In order to arouse the attention of industry stakeholders to research applications, SE researchers should ensure that others could understand why a ML/DL technique is selected for a specific SE problem and prove the replicability of their research results. Our review has shown (in Section 4.3) that it is still not a common practice for ML/DL researchers in the SE community to share their data and implementations, which makes it difficult, if not impossible, for future researchers to replicate or compare with their studies. We also noted (in Section 4.4) that most studies did not provide detailed rationales of the selected ML/DL techniques and only considered performance as the most influential factor. Without providing the details of feature preparation, the underlying assumptions and error analysis for ML/DL models, it would make it hard to reproduce methodologies in industrial settings. 
%Before moving to the next stage, we suggest to repeatedly learn from the implementation of ML/DL techniques until understanding benefits and limitation of the approaches entirely. 

The second stage primarily involves adapting ML/DL applications to fit the needs in industrial settings. This would demand close collaboration between SE researchers and industrial practitioners. Researchers should acquaint themselves with different scenarios in practice and reach a mutual agreement with practitioners on the expectation of where to improve the research applications, thus finalizing the industrial settings. After that, both parties would co-drive the process and reproduce ML/DL methodologies to fit all new changes through properly preprocessing data and optimizing the hyperparameters of ML/DL models. Nevertheless, we discovered (in Section 4.3) that SE researchers did not follow a systematic way (based on uniformed principles) to perform data preprocessing and hyperparameter tuning. This could make it hard to find appropriate data preprocessing techniques and the best combination of hyperparameters for industrial settings. It could be improved by fostering a long-term collaboration between the SE and AI communities. While successful reproduction of ML/DL applications are important, addressing quality, including scalability, reliability, efficiency, etc., is equally important. Valuable feedback can be obtained from practitioners in an iterative fashion during quality assurance until quality reaches a level that can be accepted by practitioners. Meanwhile, integrating the ML/DL process into the development process is painful and time-consuming due to many unresolved challenges \cite{Amershi:2019:SEM:3339914.3339967}, which could last until Stage 3. 

In Stage 3, SE researchers just need to support practitioners to turn applications into real-world practices. We do not claim that this road-map is the only way to generalize ML/DL applications. Instead, we would like to suggest ways to improve the state-of-the-art ML/DL-related SE research and facilitate the transfer of research results into real-world applications. The road-map also aims to stimulate the future directions for research on the synergy between ML/DL and SE, and to build a healthy eco environment for collaboration among SE/AI researchers and industrial practitioners through observing limitations from this road-map.
 
\subsection{Limitations}
One limitation of our study is the potential bias in data extraction. There were some disagreements among the three researchers with respect to determining the categories of SE tasks for several studies. To better explore the insights from the classification, the granularity of the categories should be neither be too coarse or too fine. For those questionable studies, we first identified the keywords, which indicate related tasks, in the abstract or related work according to authors' claim in order to determine their temporary categories. Then, comparing the objectives and contributions with other studies, we decided whether these temporary categories should be merged with the existing SE tasks or keep them as new tasks. The advice from the SE experts also mitigated this threat to study validity.

Another limitation might be the possibility of missing ML/DL related studies during the search and selection phases. Although it was not possible to identify and retrieve all relevant publications considering the large number of ML/DL-related SE studies, our search strategy integrated manual and automated searches together. This could keep the number of missing studies as small as possible. Finally, we acknowledged that our factors that influence replicability and reproducibility might not cover all conditions and should be continuously updated.

\section{Related Work}
\subsection{Empirical Studies for ML/DL and SE}
Some empirical and case studies of SE and ML emerged at the beginning of the 21st century. Zhang et al. \cite{Zhang2003} grouped the applications of ML methods in SE tasks based on 12 activity types and discussed the general issues of applications in the academic domain. Di Stefano and Menzies \cite{DiStefano02} suggested academic application guidelines and conducted a case study on a reuse dataset using three different machine learners. One common purpose of these studies was to stimulate more research interest, which was scarce at that time, in the areas of ML and SE. Our SLR verified their significant contribution since SE research related ML/DL applications is thriving in the past decade (2009-2018).

The most relevant study to our SLR is an industrial case study from Microsoft \cite{Amershi:2019:SEM:3339914.3339967}. They described how the Microsoft software engineering teams build software applications with ML features. The process involves integrating Agile SE processes with a nine-stage ML workflow. They also created a custom ML process maturity model for assessing the progress of building ML applications and discovered three fundamental differences in how SE applies to ML vs. previous application domains. In addition, Arpteg et al. \cite{Arpteg18} identified 12 main SE challenges specifically related to the intersection of SE practices and DL applications. In contrast, our study is more comprehensive in that it investigates not only SE for ML/DL, but also the opposite, ML/DL for SE. Nevertheless, these two studies perfectly complement our SLR, which can encourage SE researchers to address those challenges.

Many SLRs, surveys and comparative studies \cite{Zhou:2018:FWP:3208361.3183339,Tantithamthavorn17,Herbold2017,Yan17,Hall12,Shepperd14,Ghotra:2015:RIC:2818754.2818850,ARISHOLM20102} have focused on investigating the use of ML/DL in software defect prediction. Besides, Wen et al.'s \cite{WEN201241} SLR explored ML-based software development effort estimation (SDEE) models from four aspects: type of ML technique, estimation accuracy, model comparison, and estimation context. Fontana et al. \cite{ArcelliFontana2016} performed the large-scale experiment of applying 16 ML algorithms to code smells and concluded that the application of ML to detect code smells can provide a high accuracy. Compared to our SLR, the findings from the above studies are limited to several specific SE tasks and performance comparisons, rather than a comprehensive evaluation, are often the main purpose.

\subsection{Technology Transfer Model}
Technology transfer demands significant efforts to adapt a research prototype to a product-quality tool that addresses the needs of real scenarios and which is to be integrated into a mainstream product or development process \cite{Dang17}. It also requires close cooperation and collaboration between industry and academia throughout the entire research process.

Gorschek et al. \cite{Gorschek06} devised a seven-step technology transfer model to conduct an industry-relevant research in requirements engineering and product management. However, ML/DL approaches to SE differ partly from traditional SE in that their behavior is heavily dependent on data from the external world. Our road-map involves more specific steps of how to tackle both source code and data during the transfer process. Rana et al. \cite{Rana14} developed a framework for factors and attributes that contribute towards the decision of adopting ML techniques in industry for the purpose of software defect prediction.

It is worth noting that we view Figure \ref{roadmap} as a road-map that aims to narrow the gap between academia and industry and improve the eco environment for collaboration among SE researchers and industrial practitioners, rather than a mature technology transfer model or framework.

\section{Conclusion}
This paper presented a systematic literature review that comprehensively investigated and evaluated 906 state-of-the-art ML/DL-related SE studies to address four research questions that are of interest to the SE community. Our SLR showed that there was a rapid growth of research interest in the synergy between ML/DL techniques and 60 SE tasks over seven SE activities in the past decade. Through an elaborated investigation and analysis, we provided a comprehensive review of the current progress of the synergy between ML/DL and SE, which can be summarized as follows. First, we observed two positive impacts that emerging ML/DL brings to SE and the long-term benefits of ML/DL in acquring deeper insights into SE tasks. Second, we noted an absence of replicable and reproducible ML/DL applications in the SE community, which has made it difficult to generalize and transfer the research results into real-world applications. Consequently, we identified a set of five factors that may influence the replicability and reproducibility of ML/DL-related studies. Third, we noticed that many studies do not provide detailed rationales for the selection of ML/DL techniques but only consider performance as the top selection criterion. As a result, we discovered three indispensable ingredients in the studies that would facilitate the understanding of why a ML/DL technique was selected to tackle a specific SE problem. Fourth, we conducted an analysis of 79 DL studies in the SE community due to its increasingly attractiveness in SE since 2015 and the additional learning power that it has brought to SE tasks. We discussed the unique trends of impacts of DL models on SE tasks, as well as five unique challenges in integrating DL with SE, and suggested the corresponding future directions for research.
From the perspective of \textit{SE for ML/DL}, we found that it remains difficult to apply SE practices to develop ML/DL systems. Though 15 studies involved SE testing and debugging activities on ML/DL systems, they were still at an early stage. In addition, further research effort is needed to overcome the challenges involved in the integration of SE practices into more stages of the ML/DL workflow and apply them to different kinds of ML/DL systems other than supervised learning systems.

By synthesizing all the findings, we created a three-stage road-map, which suggests ways to improve the state-of-the-art ML/DL-related SE research and facilitate the transfer of research results into real-world applications. It also aims to stimulate future research on the synergy between ML/DL and SE, and to build a healthy eco environment for the collaboration between SE/AI researchers and industrial practitioners.

\bibliographystyle{unsrt}  

\bibliography{references}

\end{document}